\documentclass[preprint]{iacrtrans}
\usepackage[utf8]{inputenc}
\usepackage{cite}
\usepackage{float}
\usepackage{graphicx}
\usepackage{subfigure}
\usepackage{amssymb}
\usepackage{diagbox}
\usepackage{listings}
\usepackage{multirow}
\usepackage{soul}
\soulregister\cite7 % 针对\cite命令
\soulregister\citep7 % 针对\citep命令
\soulregister\citet7 % 针对\citet命令
\soulregister\ref7 % 针对\ref命令
\soulregister\pageref7 % 针对\pageref命令
\lstdefinelanguage
[AVR]{Assembler}     % add a "x64" dialect of Assembler
{morekeywords={mov,ori,ldi,in, %
	}} % etc.
\author[Ruizhe Gu et al.]{Ruizhe Gu, Ping Wang, Mengce Zheng, Honggang Hu, Nenghai Yu}
\institute{
	Key Laboratory of Electromagnetic Space Information, CAS\\ University of Science and Technology of China, Hefei 230026, China \email{{zheruigu,wangpin,mczheng}@mail.ustc.edu.cn}
}
\title[Adversarial Attack Based Countermeasures against Deep Learning SCA]{Adversarial Attack Based Countermeasures against Deep Learning Side-Channel Attacks}
\begin{document}

\maketitle

\keywords[Side-Channel Attacks, Countermeasures,  Adversarial Attack, Deep Learning]{Side-Channel Attacks \and  Countermeasures \and Adversarial Attack \and Deep Learning}

\begin{abstract}
  Numerous previous works have studied deep learning algorithms applied in the context of side-channel attacks, which demonstrated the ability to perform successful key recoveries. These studies show that modern cryptographic devices are increasingly threatened by side-channel attacks with the help of deep learning. However, the existing countermeasures are designed to resist classical side-channel attacks, and cannot protect cryptographic devices from deep learning based side-channel attacks. Thus, there arises a strong need for countermeasures against deep learning based side-channel attacks. Although deep learning has the high potential in solving complex problems, it is vulnerable to adversarial attacks in the form of subtle perturbations to inputs that lead a model to predict incorrectly. 
  
  In this paper, we propose a kind of novel countermeasures based on adversarial attacks that is specifically designed against deep learning based side-channel attacks. We estimate several models commonly used in deep learning based side-channel attacks to evaluate the proposed countermeasures. It shows that our approach can effectively protect cryptographic devices from deep learning based side-channel attacks in practice. In addition, our experiments show that the new countermeasures can also resist classical side-channel attacks.
\end{abstract}

%\tableofcontents{}

\section{Introduction}
Side-channel attacks (SCA) are a major threat to embedded devices \cite{yan2019attack}. They can use only a limited budget to recover the keys of cryptographic devices. The side-channel attacks exploit the side-channel information of a cryptographic computation to recover sensitive data. The side-channel information includes power consumption, electromagnetic radiations, and running-time, etc. They can recover sensitive data values in very few side-channel observations. The profiling attacks \cite{chari2002template} are one of the most powerful side-channel attacks. In this scenario, the adversary may precisely tune all the parameters of the cryptographic device, and characterize the correlation between the physical leakage and sensitive data value. They can predict the sensitive value on a target device containing a secret they wish to retrieve by using multiple traces. 

Very similar to profiling attacks, deep learning algorithms are also used in the context of side-channel attacks \cite{cagli2017convolutional, prouff2018study, kim2019make, maghrebi2016breaking, timon2019non}. Some recent studies have demonstrated the robustness of deep learning techniques to the most common countermeasures \cite{maghrebi2016breaking, masure2020comprehensive, cagli2017convolutional}. Deep learning techniques are at least as effective as classical profiled attacks. Today, security components are embedded everywhere, so deep learning based side-channel attacks have become a major threat to many everyday life objects. Facing the application of deep learning techniques in the context of SCA, the classical security protections designed to thwart classical side-channel attacks can no longer protect modern security components. Therefore, there arises a strong need for new countermeasures that can protect cryptographic devices against deep learning attacks.

\subsection{Related Work}
Singh et al. \cite{singh2018exploiting} exploited random dynamic voltage and frequency scaling to thwart SCA. Courousse et al. \cite{courousse2016runtime} presented a code morphing runtime framework to resist SCA. Boulet et al. \cite{boulet2013protection} described the protection of electronic devices against hidden-channel analysis. The protection converts the original codes to functionally equivalent codes by a modified compiler. Coron et al. \cite{coron2010analysis} mitigated side-channel attacks by the execution of dummy instructions. Ambrose et al. \cite{ambrose2007rijid} proposed to randomly insert a limited set of randomly selected instructions. They argued that such instructions could protect devices. As compared to our work, these previous works insert randomly selected instructions into the entire algorithm or the entire sensitive function. Our main contribution is to select the best suitable noise instructions and determine the exact insertion position.

Some recent studies have demonstrated the robustness of deep learning techniques against the most common countermeasures \cite{maghrebi2016breaking, masure2020comprehensive, cagli2017convolutional}. Therefore, there arises a strong need for new countermeasures that can protect cryptographic devices against deep learning attacks. In particular, to the best of our knowledge, the only former work that uses adversarial attacks to resist SCA is carried out by Picek et al. \cite{picek2019poster}. However, different from our work, they just modified each side-channel trace into adversarial samples. The experiments in this paper show that turning each side-channel trace into an adversarial trace is not an effective countermeasure.

\subsection{Our Contributions}
In this paper, we present a kind of countermeasures against the deep learning based side-channel attacks. The key idea of our approach is to add adversarial perturbations to the cryptographic algorithm implementation during compilation. We propose an approach to select adversarial perturbation instructions, and where to insert these instructions. Moreover, we also evaluate the security of our countermeasures by experiments.

In our experiments, we use two different deep learning techniques to assess the security level of our countermeasures: Multilayer Perceptron (MLP) and Convolutional Neural Network (CNN). The experimental results show that our countermeasures can reach a high level of security under deep learning based SCA. We also evaluate the performance of our approach under the classical side-channel attacks. Template attacks (TA) are exploited to attack our countermeasures, and the experiment shows that our method can thwart such attacks.

\subsection{Organization}
The paper is organized as follows. After describing notations and terminology in Section \ref{subsection-nat}, Sections \ref{subsection-sca} and \ref{subsection-aa} give some background on side-channel techniques and adversarial attacks. The threat model is described in Section \ref{section-tm}. In Section \ref{section-aabps}, we describe our countermeasures in detail. Some experiments are implemented in Section \ref{section-ee}. Section \ref{section-con} concludes this paper.

\section{Preliminaries}

\subsection{Notations and Terminology}\label{subsection-nat}
In this paper, $ k^{*} $ denotes secret keys, $\mathcal{K}$ denotes the set of all possible keys, and $\mathcal{D}_{\text {profiling}}$ denotes the profiling dataset which contains the profiling traces dataset $ \mathcal{T}_{\text{profiling}} $ and the profiling labels dataset $ \mathcal{L}_{\text{profiling}} $. The profiling traces dataset contains $ N $ profiling traces, and each trace is composed of $ n $ time samples. The profiling labels dataset contains the classes/labels which follows the ML classification meaning for each profiling trace. $\mathcal{D}_{\text {attack}}$ denotes the attack dataset, and it contains $ \mathcal{T}_{\text{attack}} $ and $ \mathcal{L}_{\text{attack}} $, where $ \mathcal{T}_{\text{attack}} $ contains $ M $ attack traces. We train neural network using $\mathcal{D}_{\text {profiling}}$ and obtain a deep learning model. $ \operatorname{Model} () $ denotes the model we trained. 

\begin{figure}[H]
	\centering 
	\includegraphics[width=8cm]{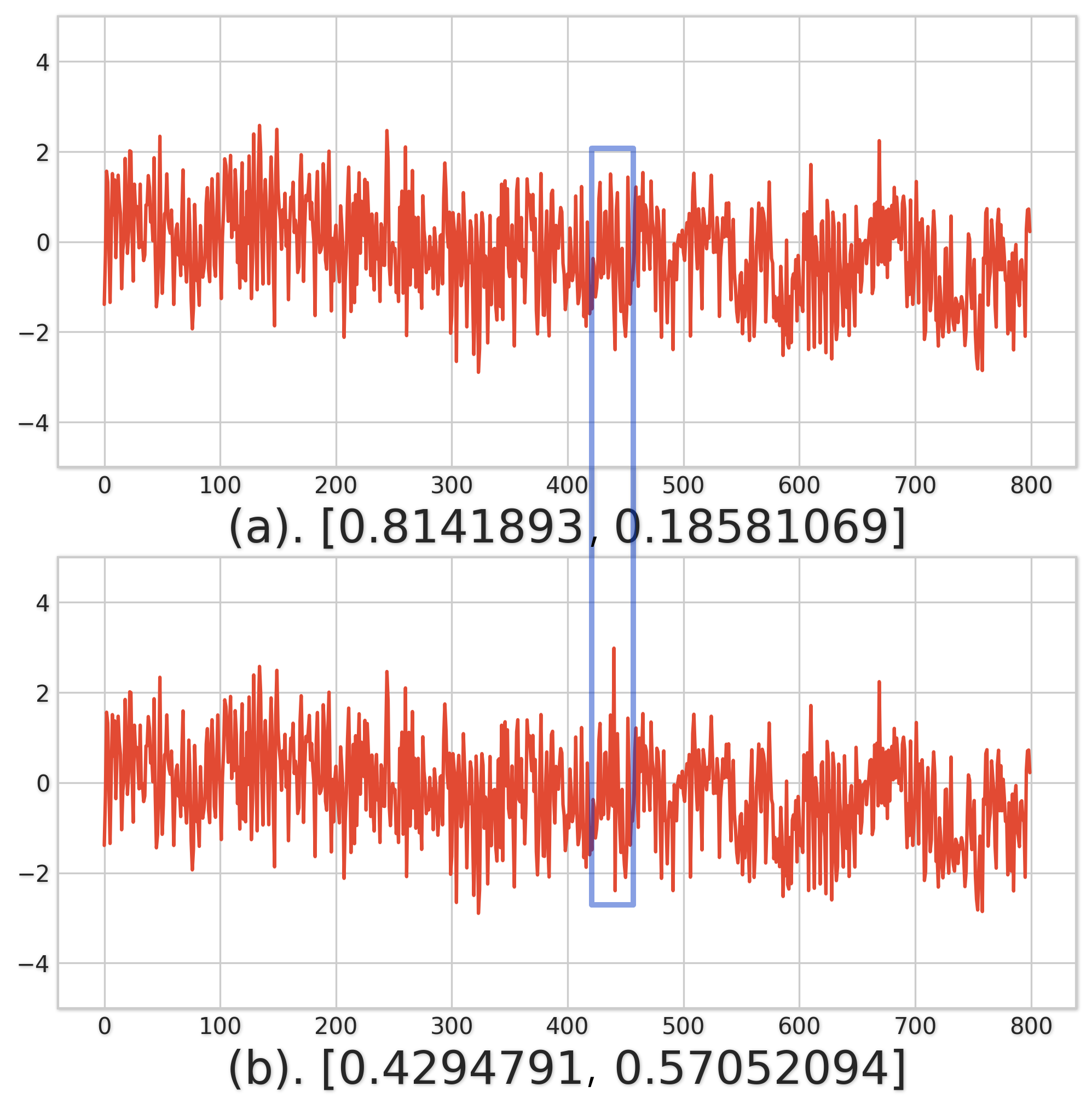}
	\caption{One-pixel attack on side-channel traces. The label of (a) is $ 0 $. The trace is labeled with the least significant bit of the output of the third Sbox during the first round, $ \operatorname{LSB}(\operatorname{Sbox}(p[3] \oplus k[3])) $. The prediction vector of (a) is $ [0.8141893, 0.18581069] $, it is classified as class $ 0 $. So the classification of (a) is correct. (b) is obtained by modifying the value of (a) at the $ 440 $th time sample. (b) is incorrectly classified as class $ 1 $.} 
	\label{Fig.onepixeltrace}
\end{figure}

Given an input trace, $ \operatorname{Model} () $ aims to compute an output called a prediction vector $ d \in \mathbb{R}^{m} $, where $m$ represents the number of possible classes/labels corresponding to the input trace. Each component in $ d$ represents the \textit{confidence} of corresponding possible class/label. For example, in Figure \ref{Fig.onepixeltrace}, the label of the traces is the least significant bit of the third key byte corresponding to the traces, $[0.4294791, 0.57052094]$ indicates that the confidence of label $0$ is $0.4294791$, and the confidence of label $1$ is $0.57052094$. The trace is classified as class $1$, because the confidence of class $1$ is greater than the confidence of class $0$.

Below we introduce some definitions, which are related to adversarial attacks against deep learning based side-channel attacks. The definitions of these terms are similar to those of adversarial attacks in computer vision area \cite{madry2018towards}. The rest of this paper follows these definitions.
\begin{itemize}	
	\item \textit{Adversarial example/trace} is a trace obtained by adding noise on the cryptographic devices to obfuscate the deep learning classifier.
	\item \textit{Adversarial perturbation} is the noise added to the cryptographic devices when generating adversarial trace.
	\item \textit{Black-box attacks} mean that the adversary attacks a deep learning model without the structure and parameters knowledge. When the adversary may have information about the structure and parameters of the model, we call it \textit{white-box attacks}.
	\item \textit{Targeted attacks} fool the deep learning models to make it misclassify adversarial traces into specified target classes/labels. They are the opposite of \textit{non-targeted attacks}. The goal of the non-targeted attack is to slightly modify the cryptographic devices in a way that the side-channel trace will be classified incorrectly by generally unknown deep learning classifier.
	\item \textit{Universal perturbation} means that the same perturbation is added to different power traces, which can make the traces misclassified by the classifier.
\end{itemize}

\subsection{Side-Channel Attacks}\label{subsection-sca}
In the real world, cryptographic algorithms always rely on a physical carrier, such as a PC, smart card, or embedded processor. When a cryptographic algorithm is running on a physical carrier, execution time \cite{kocher1996timing}, power consumption \cite{kocher1999differential}, electromagnetic emissions \cite{gandolfi2001electromagnetic}, and other side-channel information of a cryptographic computation are leaked. These side-channel leakages of a cryptographic computation depend on some small part of the internally used sensitive data or sensitive operations in the cryptographic devices, and can be exploited to recover keys. A key-recovery attack based on side-channel leakage analysis is called a side-channel attack for simplicity.

\subsubsection{Template Attack}
\label{Section.2.2.2}
TA can be considered as the most successful and in-depth research method in classical SCA. In this paper, we use TA to evaluate the security level of our countermeasures. Let us consider the target device executing a cryptographic algorithm with the secret key $ k^{*} $. The adversary may control a copy of the target device called profiling device and priorly use it to precisely tune all the parameters of the cryptographic computation. For each possible key $ k $ the adversary observes $ N^{(k)} $ time over a time interval of $ n $ time samples the power consumption of profiling device and we denote by \textit{trace} the series of observations $ T_{(i)}^{(k)}=\left\{T_{(i)(t)}^{(k)} \in \mathbb{R} | t \in[1 ; n]\right\}, i=1, \ldots, N^{(k)} $. The most common TA model modelizes the stochastic dependency between $ k $ and \textit{trace} by means of a multivariate normal conditional density:
\begin{equation}
P\left(T_{(i)}^{(k)} | k \right)=\frac{1}{\sqrt{(2 \pi)^{n}\left|\Sigma_{k}\right|}} e^{-\frac{1}{2}\left(T_{(i)}^{(k)}-\mu_{k}\right) \Sigma_{k}^{-1}\left(T_{(i)}^{(k)}-\mu_{k}\right)^{\prime}}
\label{pipeigailv1}
\end{equation}
where $\mu_{k} \in \mathbb{R}^{n}$ and $\Sigma_{k} \in \mathbb{R}^{n \times n}$ are the expected value and the covariance of the $ n $ variate traces respectively.

In the context of TA, two phases may be distinguished: 

\paragraph{Profiling Phase.}
For each possible key $ k $, the adversary captures $ N^{(k)} $ traces $ T_{(i)}^{(k)} $ over a time interval of length $ n $. TA estimates the expected value $\mu_{k}$ and the covariance $\Sigma_{k}$ by
 
\begin{equation}
\hat{\mu}_{k}=\frac{1}{N^{(k)}} \sum_{i=1}^{N^{(k)}} T_{(i)}^{(k)}
\label{junzhimuban}
\end{equation}

\begin{equation}
\hat{\Sigma}_{k}=\frac{1}{N^{(k)}-1} \sum_{i=1}^{N^{(k)}}\left(T_{(i)}^{(k)}-\hat{\mu}_{k}\right)^{\top}\left(T_{(i)}^{(k)}-\hat{\mu}_{k}\right)
\label{zaoshengmuban}
\end{equation}

\paragraph{Attack Phase.}
The attacker captures a trace $ T $ when the target device execute a cryptographic algorithm. The adversary estimates the secret key which maximizes the likelihood:

\begin{equation}
\hat{k}=\arg \max _{k} P\left(T | k \right)
\label{equ_secretkey}
\end{equation}

\subsubsection{Deep Learning Based SCA}\label{subsection-dl}
Deep learning based side-channel attacks focus mainly on two techniques: multi-layer perceptrons (MLP) and convolutional neural networks (CNN) \cite{cagli2017convolutional}. Martinasek et al. \cite{martinasek2016profiling,martinasek2013optimization,martinasek2015profiling} compared MLP-based methods with other classical attack such as template attacks. Cagli et al. \cite{cagli2017convolutional} have shown that MLP-based attack is far more effective than other classical methods. Prouff et al. \cite{prouff2018study} have demonstrated that CNN can obtain a great success in attacking cryptographic implementations with jitter.

Deep learning based SCA \cite{maghrebi2016breaking} is similar to TA, but uses deep learning techniques as a profiling method instead of using multivariate Gaussian profiling as in TA. To train a deep learning model, the typical leakage models used for the power consumption are the Hamming Weight (HW) model ($ 9 $-class classification), and the Least Significant Bit (LSB) model ($ 2 $-class classification) \cite{perin2018lowering}. In this paper, we also uses these two leakage models.

\paragraph{MLP.}
 MLP is also called artificial neural networks. It contains at least three layers: in addition to the input and output layers, there can be multiple dense layers between them. The number of neurons in the input layer is determined by the number of time samples $ n $ in the input data. The MLP layer is fully connected (fully connected means that each neuron in the upper layer is connected to all neurons in lower layer). If the output of the lower layer is represented by a vector X, the output of the higher layer is $ f\left(w_{i j} x+b_{i j}\right) $, where $ w_{i j} $ and $ b_{i j} $ are the weight and bias of the $ j $-th neuron in the $ i $-th layer respectively, and $ x \in X $. Generally, the function $ f $ is $ sigmoid $ or $ tanh $. Finally, the output layer can be viewed as multi-class logistic regression, i.e. softmax regression. Thus, the output of the output layer is $ softmax(wx + b) $. The parameters of the MLP are all the connection weight $ w $ and the bias $ b $ between the layers. The process of training these deep learning models is the process of finding the optimal parameters. How to set the optimal parameters is an optimization problem. To solve the optimization problem, the easiest method is the gradient descent method.

\paragraph{CNN.}
 CNN can be regarded as a variant of MLP. In addition to the input layer, dense layer and output layer, it also uses one or more convolutional layers and pooling layer. A convolution layer includes a convolution operation, followed by an activation function (such as ReLU) and a pooling layer. The pooling layer is used to reduce the dimensions. The convolutional layer performs a series of convolutional operations on its inputs (each input is convoluted with a filter).

\subsection{Adversarial Attacks}\label{subsection-aa}
The adversary may design a targeted machine learning sample (\textit{adversarial example/trace}) to make the machine learning model misjudge. This is called an adversarial attack. Szegedy et al. \cite{szegedy2013intriguing} first discovered an interesting weakness of deep neural networks in image classification. Their study shows that, despite the high accuracy of deep learning, it is surprising that they are susceptible to adversarial attacks in the form of applying small \textit{adversarial perturbations} on source images. Mohse et al. \cite{moosavi2017universal} demonstrated the existence of \textit{universal perturbations}, which can be used to fool deep learning classifiers by adding it to arbitrary images. This work inspires us to protect cryptographic devices by adding \textit{universal perturbations}. 

In fact, the deep learning based side-channel attacks are also classification problems. They use deep learning techniques to classify side-channel traces. The labels for the side-channel traces are key-related values. Adversarial attacks can be seen as the process of seeking a vector $ v $ such that:
$$\operatorname{Model}(T+v) \neq \operatorname{Model}(T) $$
side-channel tarce $ T \in \mathcal{T} $. For each trace $ T $, Model() outputs an estimated label $ \operatorname{Model}(T) $. $ v $ is an \textit{adversarial perturbation}. In the context of image classification, in order to make the \textit{adversarial perturbation} less perceptible, $ v $ is often restricted to satisfy certain restrictions. If there is a $ v $ such that
$$\operatorname{Model}(T+v) \neq \operatorname{Model}(T) \text { for ``most" } T \in \mathcal{T},$$
then $ v $ is \textit{universal perturbation}.

These universal perturbations are not only universal across side-channel traces, but also generalize well across deep learning models \cite{szegedy2013intriguing,kurakin2016adversarial,moosavi2017universal}. The deep learning models find the decision boundaries of the data in the high-dimensional space. In order to make $ v $ as small as possible, the adversarial perturbations are all in the neighbourhood of decision boundaries. Even different models are used to classify side-channel traces, as long as the models are efficient, the decision boundaries they find are similar. Such perturbations are therefore doubly universal, both with respect to the data and the models. That is, if we use one model to generate a set of universal perturbations, we can find that these perturbations are still effective for another model even it was trained with different hyperparameters or it is trained on a different set of traces.

\subsubsection{One-Pixel Attack}
One-pixel attack, a type of adversarial attack techniques, which fools the deep learning classifier by changing only one pixel in the image. In order to reduce the number of inserted noise instructions, we hope our countermeasures to modify as few pixels as possible. Thus, in this paper we use one-pixel attack to calculate \textit{universal perturbations}.

Su et al. \cite{su2019one} claim to achieve an extreme case in adversarial attacks, and they fool the deep learning classifier by changing only one pixel in the image. One-pixel attack is also effective on the side-channel traces, and we show in Figure \ref{Fig.onepixeltrace} an \textit{adversarial sample} on the side-channel traces generated by one-pixel attack. The upper trace is the original trace captured during the cryptographic computation. The label for the original trace (a) is $ 0 $, and the prediction vector calculated by classifier is $ \left[ 0.8141893, 0.18581069\right] $. The deep learning classifier can correctly classify the trace (a). We use one-pixel attack to generate an adversarial trace (b) based on the trace (a). Trace (b) is obtained by changing one time sample in trace (a). The modified time sample is highlighted with blue rectangle. We use classifier to calculate the prediction vector of (b): $ \left[ 0.4294791, 0.57052094\right] $. Trace (b) is incorrectly classified as class 1 by the deep learning classifier.

One-pixel attack generates the adversarial samples using differential evolution algorithm \cite{das2010differential}. Differential evolution (DE) is a population based optimization algorithm for solving complex multi-modal optimization problems\cite{das2010differential, storn1997differential}. The differential evolution algorithm is composed of three phases: \textit{mutation}, \textit{crossover}, and \textit{selection}. Mutation is a method used to generate random solutions. Crossover is used to enhance the diversity of random solutions. Selection removes solutions that fail to evolve, and leaves solutions that succeed in evolution. The differential evolution algorithm flow is as follows: first, a sufficient number of random variables are generated as the initial possible solution. Then, the mutation, crossover, and selection are performed in order. After completing a round, a certain termination condition is checked. If the termination conditions have not been met, the differential evolution algorithm returns to mutation, crossover, and selection; otherwise, the algorithm terminates, and outputs the best solution of the last round.

When we use adversarial attack to protect cryptographic devices, the less adversarial perturbations we insert, the easier our countermeasures can be implemented. For one-pixel attack, we only need to add noise at one time sample. Besides, one-pixel attack requires less network information as it is a black-box attack.

\section{Threat Model}\label{sectionTM}\label{section-tm}
The adversary targets the secret key $ k^{*} $ of a cryptographic device. We call this cryptographic device the target device. The adversary has the same device as the target device, called the profiling device. We consider that the cryptographic device has sufficient computational resources to compile the code before each encryption (our countermeasures insert noise instructions during compilation), and the adversary cannot get control over the code compilation.

We also assume that the adversary does not use any preprocessing techniques on power traces, but we argue that the preprocessing algorithm cannot break our countermeasures. The effect of preprocessing techniques on our countermeasures is discussed further in Section \ref{section-con}. The adversary uses these two devices to carry out deep learning based side-channel attacks. The deep learning based SCA is divided into two phases:

\paragraph{Profiling Phase.}
The attacks are performed on the $ 3 $rd key byte of the AES-128 , which is the same as previous work \cite{prouff2018study}. In this case, we can refer to this previous work to obtain an effective deep learning model. For each key candidate value $ k $, the adversary captures $ N^{(k)} $ power traces. All these traces make up profiling traces dataset $ \mathcal{T}_{profiling} $. In order to analyze the effectiveness of our countermeasures on deep learning models with different output classes, the side-channel traces are labeled using two leakage models: LSB and HW. The adversary trains neural network using $ \mathcal{L}_{profiling} $ and $ \mathcal{T}_{profiling} $, and obtain a deep learning model $ \operatorname{Model} () $.

\paragraph{Attack Phase.}
The adversary captures $ M $ side-channel traces on the target device. For each $ T_{(i)} \in T_{attack} $, the adversary uses the deep learning model $ \operatorname{Model} () $ to get a prediction vector $ d_{(i)} \in \mathbb{R}^{|\mathcal{K}|} $: 
\begin{equation}
d_{(i)}=\operatorname{Model}\left(T_{(i)}\right). 
\label{zhixinjuzhen}
\end{equation}
The adversary selects the key candidate with the highest sum confidence as the secret key $ k $, i.e. $ k =\operatorname{argmax}_{k \in \mathcal{K}}\left(\prod _{i=1}^{M} d_{(i)}\right)[k] $. If $ k $ = $ k^{*} $, the key recovery is successful.

\section{Adversarial Attack Based Security Protections}\label{section-aabps}

\subsection{Differences}
\label{subsection.dbicadlaca}
One-pixel attacks were first proposed to attack deep-learning models in the image classification area. Although both deep learning based image classification and deep learning based side-channel attacks use deep learning models as  classifiers, there are still many differences between them. 

\paragraph{Different training sets.}

\begin{figure}[H]
	\centering 
	\includegraphics[width=8cm]{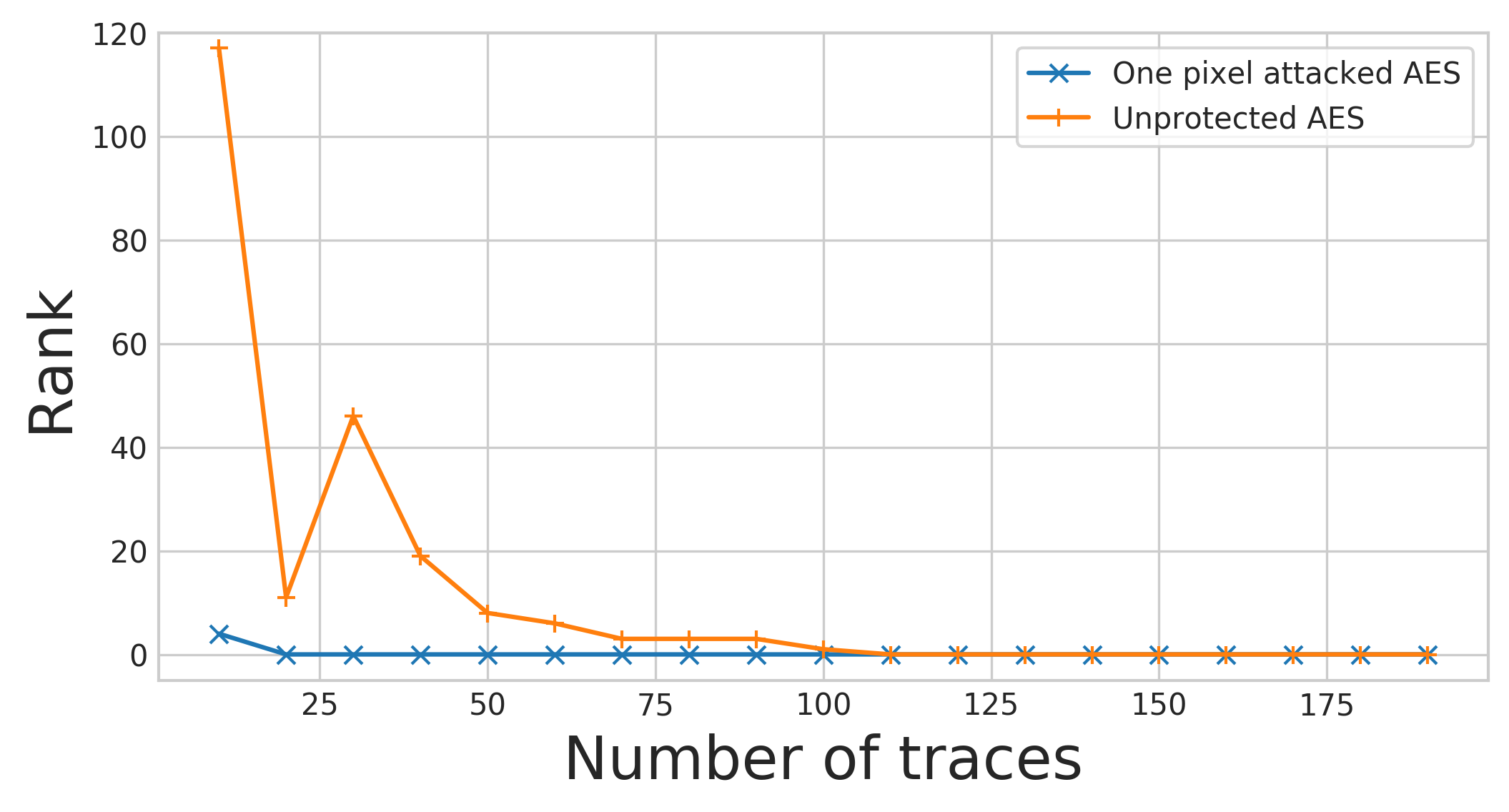}
	\caption{The mean \textit{rank} of the unprotected AES and the one-pixel attacked AES on CNN-based attack. The \textit{rank} is a metric to evaluate the security level of countermeasures (described in Section \ref{subsectionEM}). For unprotected AES, approximately $ 100 $ traces are required for a full success of the key recovery. For one-pixel attacked AES, performing successful key recoveries requires approximately $ 20 $ traces.} 
	\label{Fig.rank}
\end{figure}

In the context of image classification, we train the deep learning model on source images. However, for the deep learning based SCA, the profiling traces are captured on the profiling device. In our threat model, the profiling device is a copy of target device and it is a cryptographic implementation with our countermeasures. Therefore, if our countermeasures modify each trace to adversarial trace by adding adversarial perturbations, then the profiling traces captured on profiling device are all adversarial traces. Due to the robustness of deep learning technique, such countermeasures that insert perturbations to turn side-channel traces into adversarial traces cannot protect cryptographic devices. The attack results shown in Figure \ref{Fig.rank} confirm this view.

We first collect $ 60,000 $ power traces of unprotected AES, and we call these $ 60,000 $ traces \textit{source traces}. $ 50,000 $ source traces are used as the training set, and $ 10,000 $ source traces are used as the test set, and CNN is used to attack these source traces. The attack result is shown as unprotected AES in Figure \ref{Fig.rank}. We use the one-pixel attack to generate $ 60,000 $ adversarial traces, and we call these adversarial traces as one-pixel attacked AES traces. We use $ 50,000 $ one-pixel attacked AES traces as training set and $ 10,000 $ as test set, and use CNN to attack these traces. The attack result is shown as one-pixel attacked AES in Figure \ref{Fig.rank}. For unprotected AES, approximately $ 100 $ traces are required for a full success of the key recovery. For one-pixel attacked AES, performing successful key recoveries requires approximately $ 20 $ traces. Figure \ref{Fig.rank} shows that converting the source traces to adversarial traces cannot protect the cryptographic devices, but makes the implementation more vulnerable. The reason is that these adversarial traces have high confidence in the wrong label, which is generally above $ 0.95 $. These adversarial traces can deceive models trained on source traces. However, when adversary trains models on these adversarial traces, these adversarial perturbations which fool the original model will instead become features exploited by the deep learning techniques.

\paragraph{Different attack dataset sizes.}
In the image classification area, the purpose of the adversarial attack is to make a certain image misclassified by a deep learning model after adding perturbations that is not perceived by humans. We considered it a successful adversarial attack if the image was misclassified after adding perturbations. Therefore, for image classification area, the size of the attack dataset can be regarded as $ 1 $. In SCA area, to perform successful key recoveries, the adversary captures $ M $ power traces on the target device and selects the key candidate with the highest sum confidence as the secret key, as mentioned in Section \ref{sectionTM}. The size of the attack set can be regarded as $ M $. In order to thwart the deep learning SCA, our countermeasures need to modify all the power traces generated by cryptographic computations instead of modifying one trace.

\subsection{Our Method}
Our countermeasures insert noise instructions into the code. The power consumption of these noise instructions becomes universal perturbations. These universal perturbations make power traces misclassified by the deep learning models, and then thwart the deep learning SCA. In this process, we need to solve three problems: how to determine the position where the noise instruction is inserted, which instructions are inserted into the code as noise instructions, and how to insert the noise instruction at the selected position. We address these issues in the following subsections.

\subsection{Locations of Noise Insertion}
\label{Section.Loni}
We want to insert noise instructions at the locations where the \textit{universal perturbations} are located. Therefore, we generate \textit{universal perturbations}, and observe their positions on power traces. Before calculating the \textit{universal perturbations}, we need to determine what kind of \textit{universal perturbations} we need to calculate. Different \textit{universal perturbations} have different effects on the deep learning classifiers. Some universal perturbations make the confidence of a certain class very large, but some make the confidence of a certain class very small. These effects depend on the termination condition of adversary attacks.

This subsection calculates the locations where the \textit{universal perturbations} are located based on the $ 2 $-class model (the traces are labeled as LSB of sensitive value). We analyze at the end of this subsection that the positions calculated by 2-class model and $ 9 $-class model are close. We consider the formula $ k =\operatorname{argmax}_{k \in \mathcal{K}}\left(\prod _{i=1}^{M} d_{(i)}\right)[k] $, used by the adversary to recover the secret key. The adversary selects the key candidate with the highest sum confidence as the secret key. To prevent the adversary from recovering the correct key, we can make the confidence corresponding to a certain class of all side-channel traces very large or make the confidence corresponding to each class the same. In this way, the adversary cannot recover the correct key. We test two termination conditions: the first termination condition is that the algorithm terminates when each trace is classified as $ 0 $, and make the confidence of label $ 0 $ the as large as possible, i.e. $ d_{k=0} \geqslant \tau $, where $ d_{k=0} $ denotes the confidence corresponding to label $ 0 $ and $ \tau $ is a constant close to $ 1 $. Another termination condition: the algorithm terminates when the difference between the confidence of label $ 0 $ and label $ 1 $ within a small range, i.e. $ \left|d_{k=0}-d_{k=1}\right| \leqslant \sigma $, where $ \sigma $ is a constant close to $ 0 $. 

\begin{figure}[H]
	\centering
	\subfigure[2-class MLP model.]{
		\includegraphics[width=6.5cm]{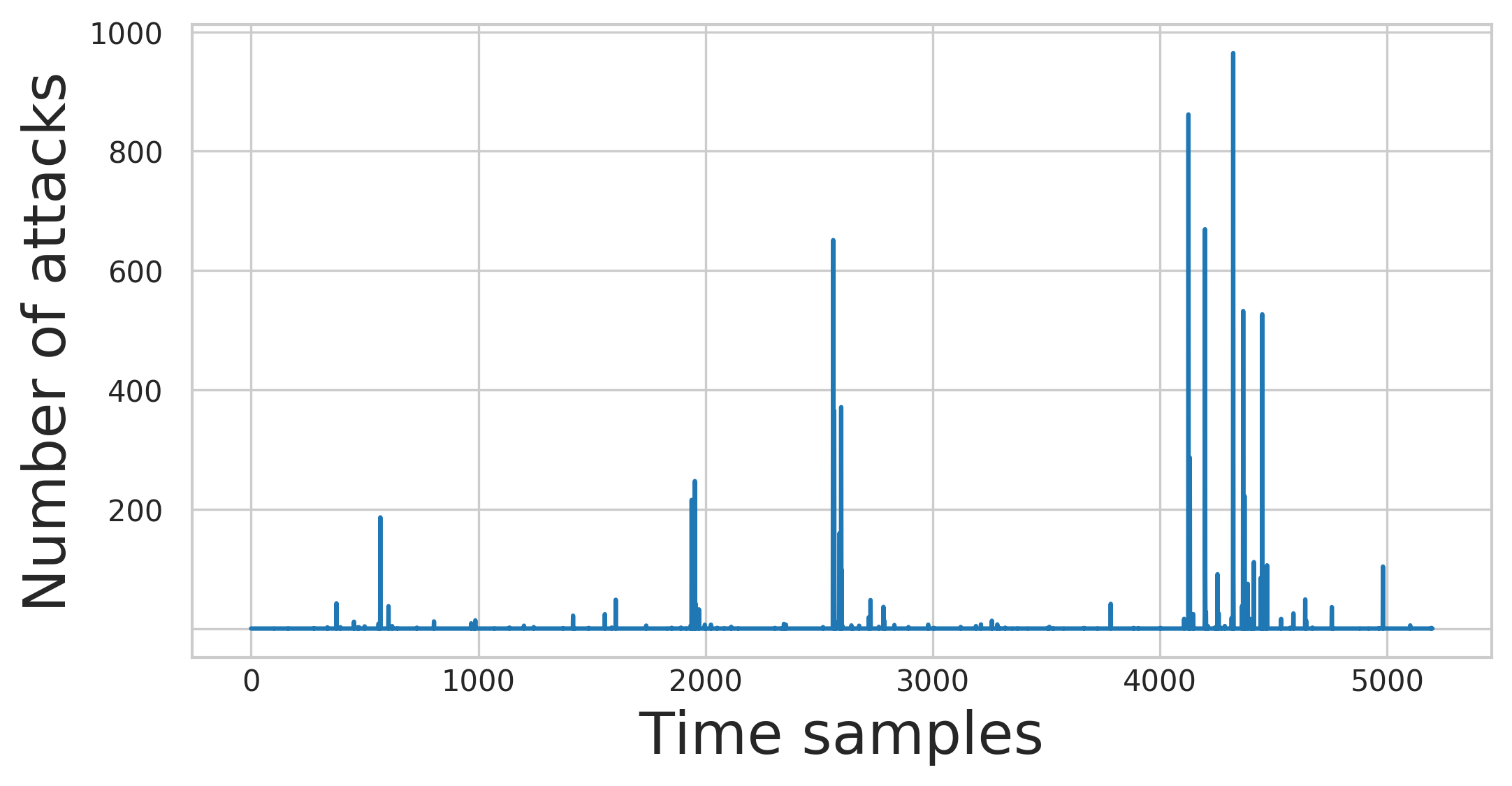}
		\label{Fig.MLPopa}
	}
	\subfigure[ 2-class CNN model.]{
		\includegraphics[width=6.5cm]{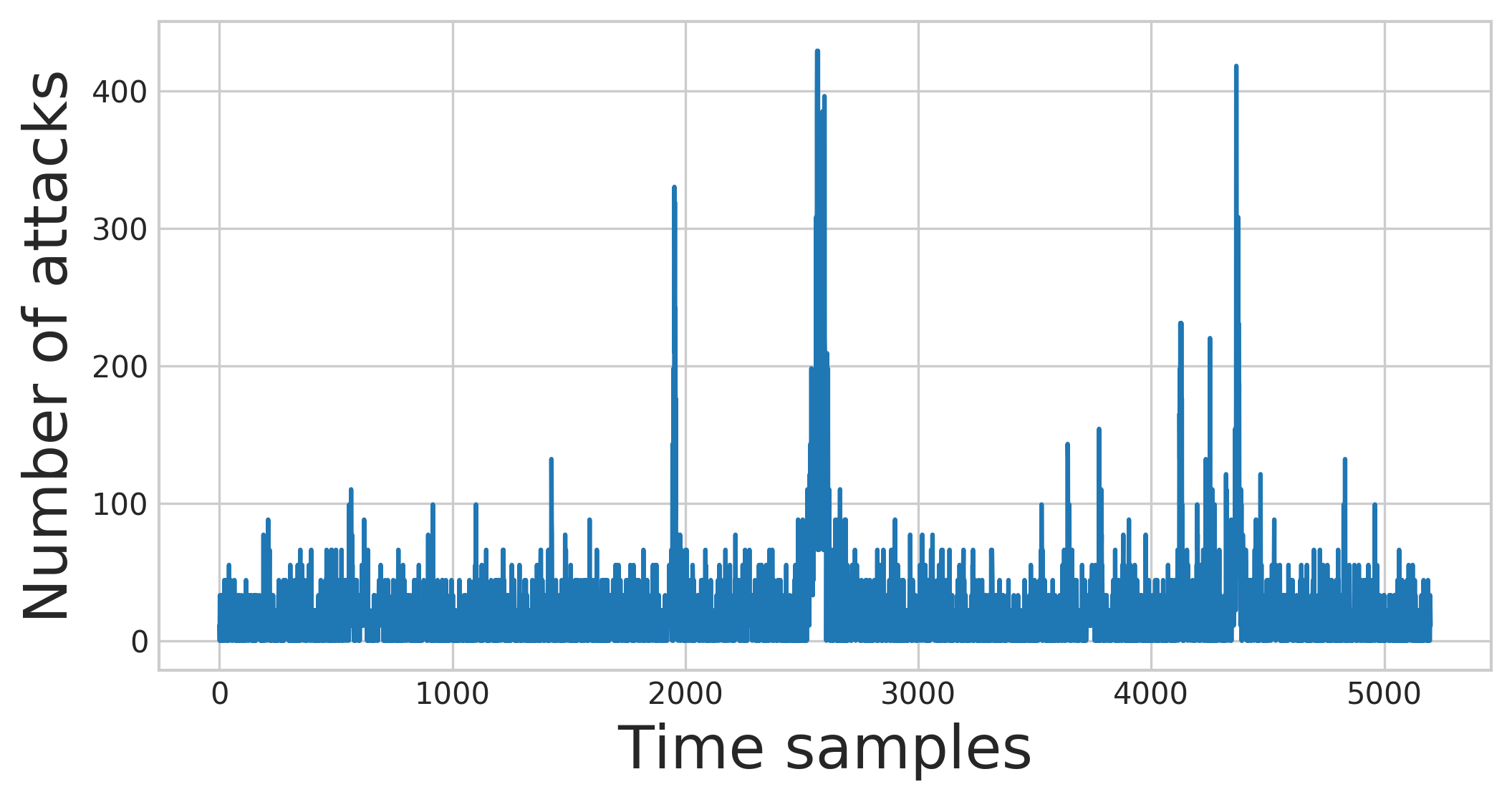}
		\label{Fig.CNNopa}
	}
	\subfigure[9-class MLP model.]{
		\includegraphics[width=6.5cm]{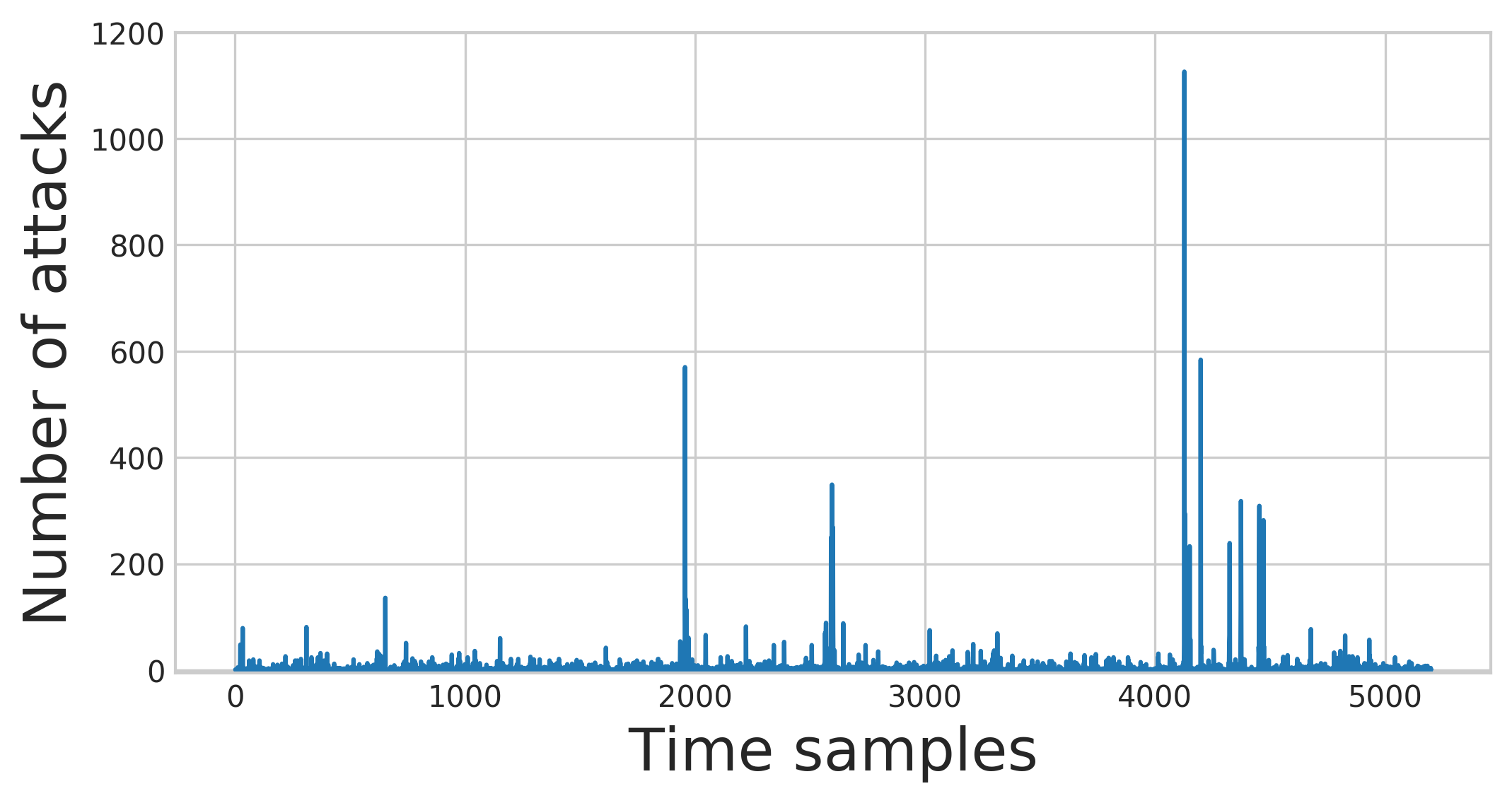}
		\label{Fig.2classMLPopa}
	}
	\caption{Distribution of adversarial perturbation. The horizontal axis represents $ 5200 $ time samples of the side-channel trace, and the vertical axis represents the number of adversarial perturbations falling on a certain time sample.}

\end{figure}

Our experiments find that the second termination condition is computationally intensive when running differential evolution algorithms. Moreover, when the leakage model is not LSB, the labels of the traces are no longer only $ 0 $ and $ 1 $, and the second termination condition is hard to implement. Thus, considering efficiency and versatility, our countermeasures use the first termination condition. The steps of calculating the insertion position are as follows:
\begin{enumerate}
	\item We capture $ 60,000 $ labelled traces of the power consumption of unprotected AES implementation. The temporal acquisition window is set to record the first round of the AES only. Each trace is composed of $ 5200 $ time samples. We experimentally validate that the deep learning classifier trained on $ 50,000 $ profiling traces can successfully recover the key in less than $ 1000 $ attack traces. Therefore, we select $ 50,000 $ traces out of $ 60,000 $ traces as profiling traces, and train deep learning models on these profiling traces.
	\item Differential evolution algorithm is applied to generate adversarial perturbations based on the remaining $ 10,000 $ traces. The termination condition is that the algorithm terminates when the trace is classified as $ 0 $. 
	\item We generate $ 10,000 $ adversarial perturbations based on MLP and CNN respectively. The distribution of the $ 10,000 $ adversarial perturbations on $ 5200 $ time samples is shown in Figures \ref{Fig.MLPopa} and \ref{Fig.CNNopa}. Figures \ref{Fig.MLPopa} and \ref{Fig.CNNopa} show that when attacking CNN and MLP, the distributions of adversarial perturbation are similar. They all have the largest distribution around the three time samples: $ 1900 $th time sample, $ 2560 $th time sample, and $ 4300 $th time sample. The perturbations near these three positions are \textit{universal perturbations}.
	\item We use these three points as the locations of the noise insertion. Noise instructions inserted near these three time samples can generate universal perturbations.
\end{enumerate}

It can be observed in the power traces of AES that these three time samples are respectively included in three functions: AddRoundKey(), SubBytes() and MixColumns(). These functions may contain thousands of instructions, and we need to know where these time samples are in the cryptographic code more accurately. Only by knowing the specific location of the noise in the code can we accurately insert noise instructions into the code. 

The C file is compiled to an assembly code file. We use the binary search algorithm to traverse the instructions of the assembly file, and insert the trigger\_low() function after the instruction. This function is also the trigger signal used when we collect the power traces. Then we compile and run the file again, and observe whether the power trace becomes low level near the three time samples we selected earlier. If it becomes low level, then the position of trigger\_low() is the position where we want to insert the noise instructions. If not, we continue to traverse the assembly file, and repeat the previous steps. In the process of determining the position of insertion, we find that even if the position of inserting the low-level signal is the same, the position of the low-level on the power trace is not the same, but their positions are very close on the power trace. The universal perturbations we generated are not all at a certain time sample but are concentrated around those three samples. Our experiments also find that if a perturbation with side-channel profile which is similar to the side-channel profile of universal perturbation is inserted near those three samples, the deep learning model can be deceived. Our purpose is only to find a fuzzy location, so that the side-channel leakage generated by the noise instructions is near to those three sampling points.

We use the $ 9 $-class MLP model to generate the adversarial perturbations, and observe their distribution on the time sample. The results are shown in Figure \ref{Fig.2classMLPopa}. The distribution of these perturbations is similar to the distribution of perturbations generated by the $ 2 $-class model. Since universal perturbations are doubly universal, even if we use different deep learning models to generate universal perturbations, the positions of these universal perturbations are close.

\subsection{Choice of Noise Instructions}
Previous works \cite{courousse2016runtime,belleville2018automated,amarilli2011can,agosta2019compiler} insert noise instructions between each useful instruction in sensitive function Sbox. These inserted noise instructions are commonly used in cryptographic algorithms. However, the purpose of our noise insertion is different from these previous literatures. In previous works, the purpose of inserting noise is to move the point of information leakage in time and space, and to reduce side-channel leakage. Thus, in these previous literatures, the inserted noise instructions do not need to be carefully selected. The purpose of our noise instruction is to make the captured side-channel traces into adversarial examples. 

So the instructions need to meet the following requirements: 1) the side-channel profile (i.e. power consumption or electromagnetic radiation) of noise instructions should be as close as possible to the profile of useful instructions, so that the adversary cannot distinguish them and filter them out from the side-channel traces \cite{durvaux2012efficient}; 2) the side-channel profile of the inserted instructions should be similar to the profile of the adversarial perturbations. The first requirement is easy to achieve, we only need to choose the instructions commonly used in cryptographic algorithms, such as addition, subtraction, exclusive or, and load. In order to meet the second requirement, the side-channel profile of adversarial perturbations should be taken into account. We analyze the distribution of adversarial perturbations over amplitude.

\begin{figure}[h]
	\centering
	\subfigure[Amplitude distribution of perturbations near the $ 1900 $th time sample]{
		\includegraphics[width=10cm]{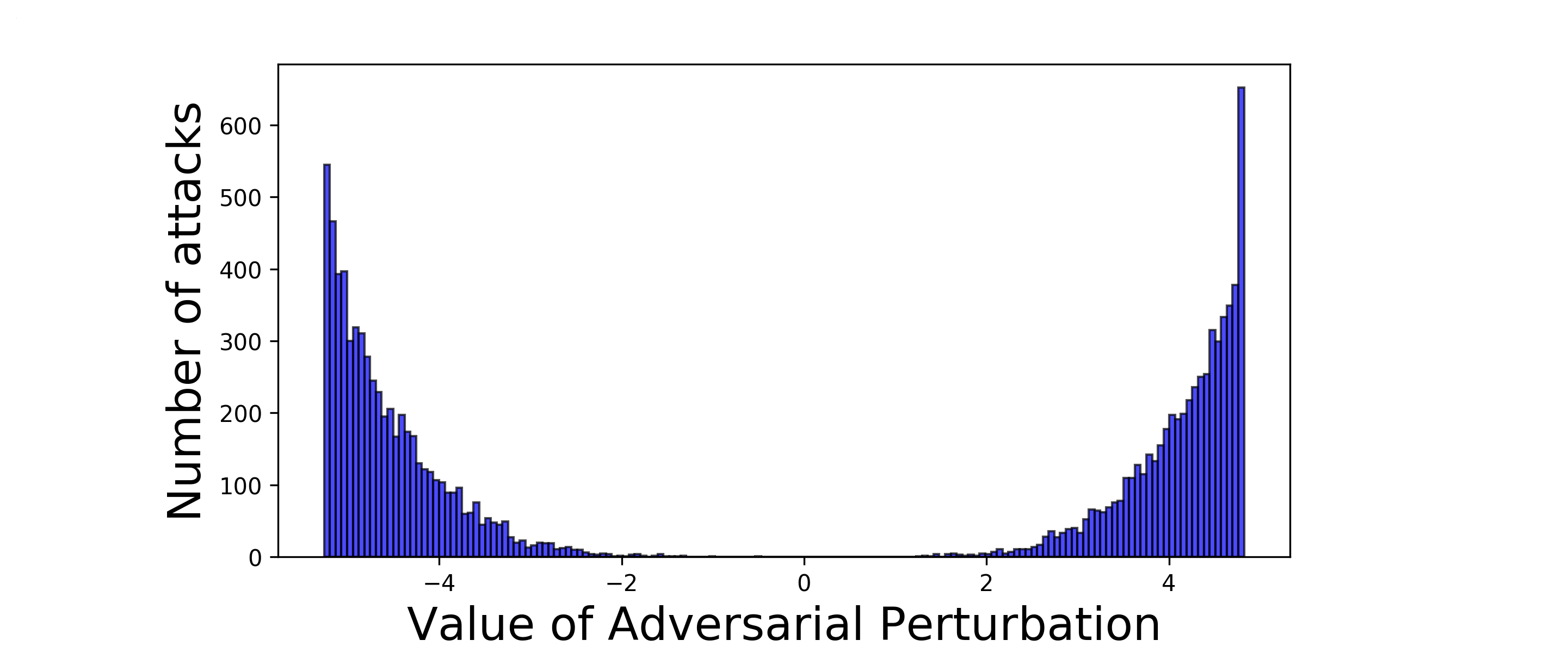}
	}
	\subfigure[Amplitude distribution of perturbations near the $ 2560 $th time sample]{
		\includegraphics[width=10cm]{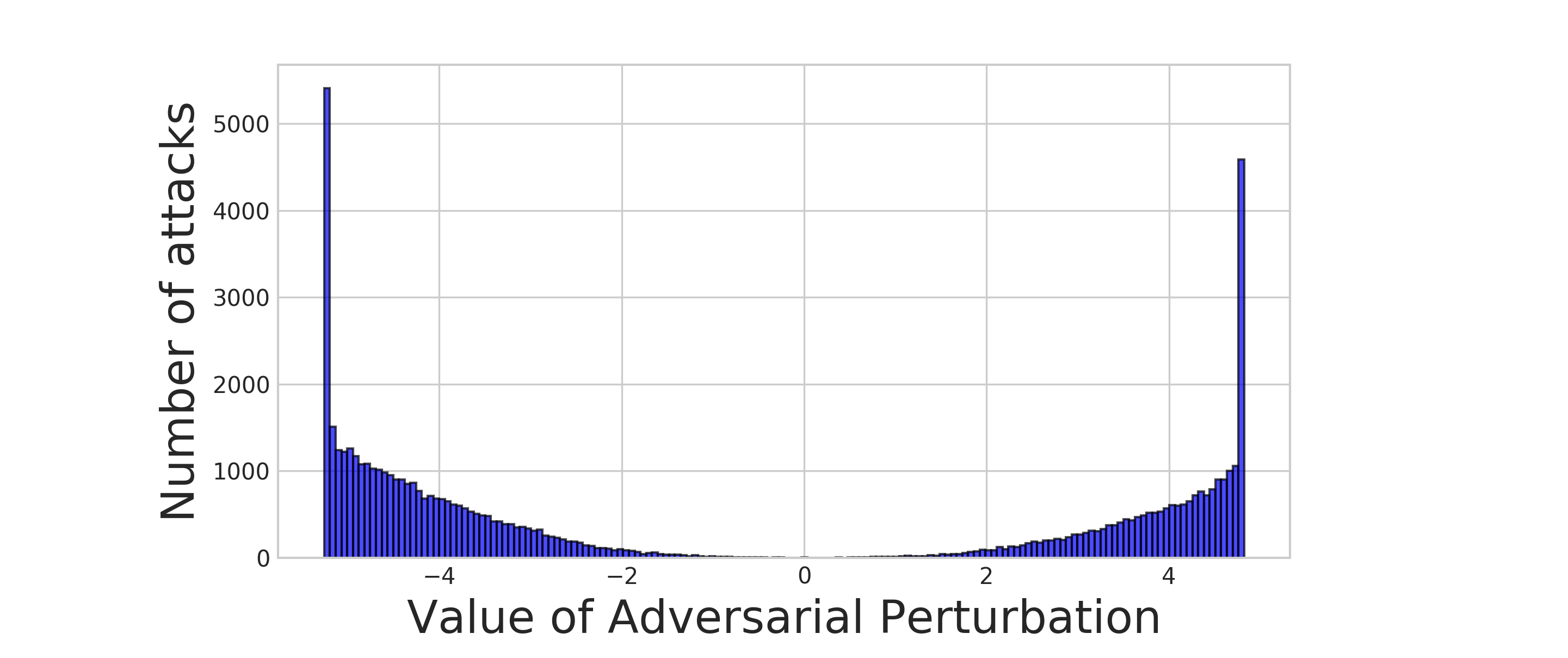}
	}
	\caption{Amplitude distribution of adversarial perturbation on MLP model. The values on the horizontal axis corresponds to the amplitude of power traces. The above picture shows the amplitude distribution of perturbations near the $ 1900 $th time sample, and the below picture shows the amplitude distribution of perturbations near the $ 2560 $th time sample. We divide the interval of the amplitude $ [-5.2, 4.8] $ into $ 160 $ discrete intervals.}
	\label{Fig.MLP-ValueDistr}
\end{figure}

\begin{figure}[h]
	\centering
	\subfigure[Amplitude distribution of perturbations near the $ 1900 $th time sample]{
		\includegraphics[width=10cm]{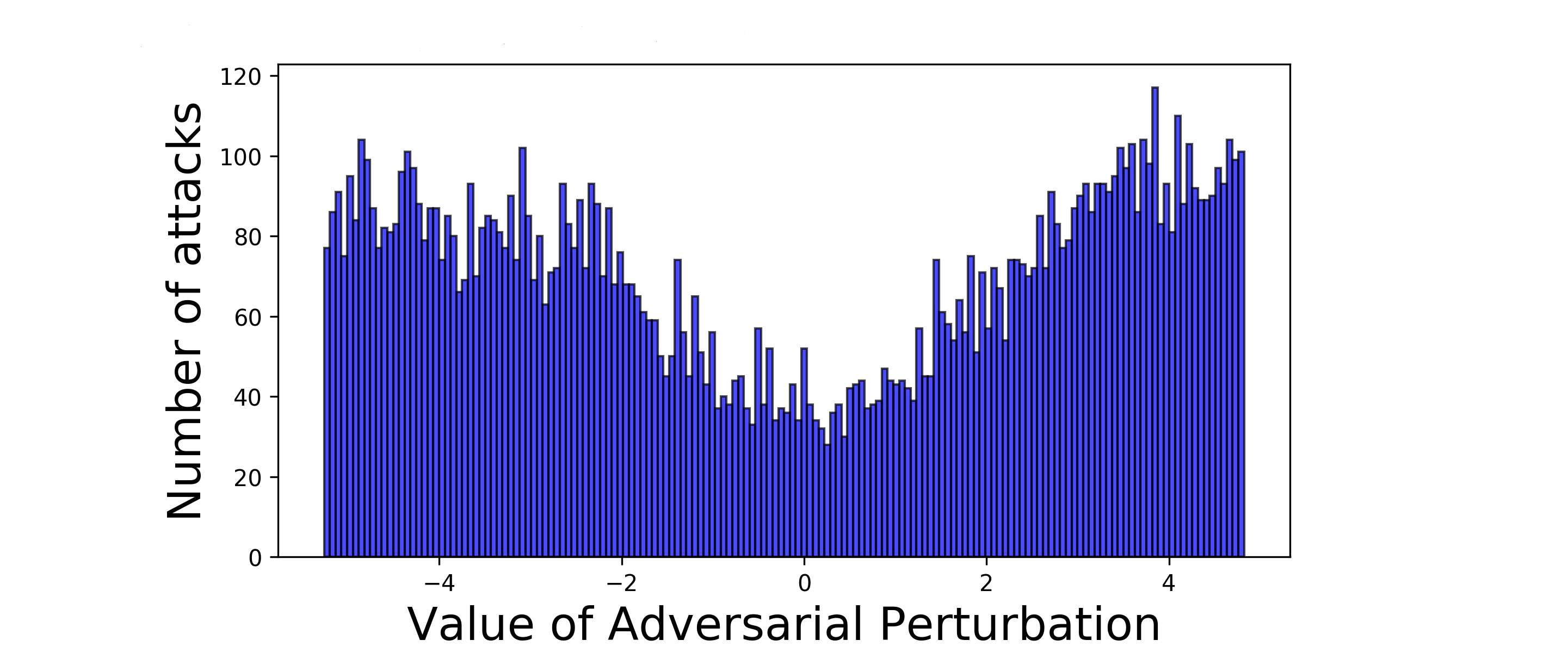}
	}
	\subfigure[Amplitude distribution of perturbations near the $ 2560 $th time sample]{
		\includegraphics[width=10cm]{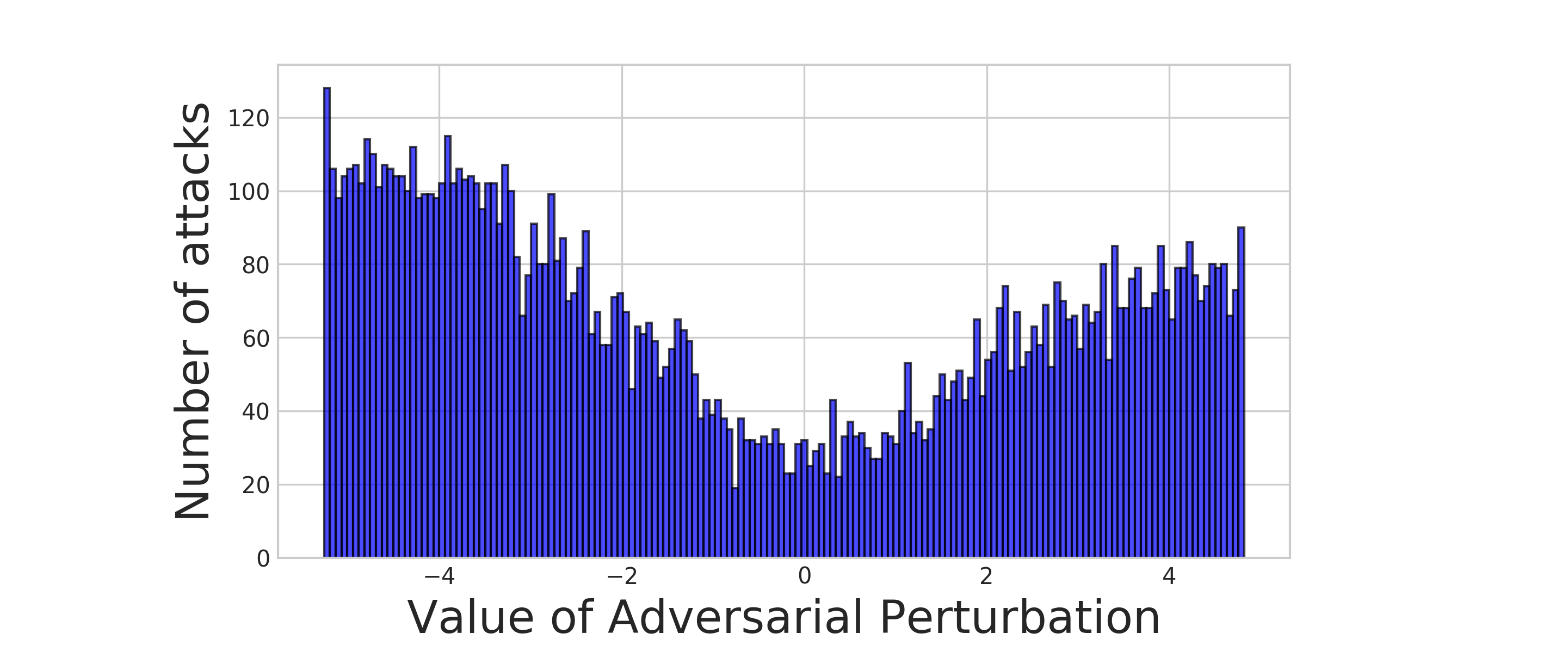}
	}
	\caption{Amplitude distribution of adversarial perturbation on CNN model. The values on the horizontal axis corresponds to the amplitude of power traces. The above picture shows the amplitude distribution of perturbations near the $ 1900 $th time samples, and the below picture shows the amplitude distribution of perturbations near the $ 2560 $th time samples. We divide the interval of the amplitude $[-5.2, 4.8]$ into $ 160 $ discrete intervals.}
	\label{Fig.CNN-ValueDistr}
\end{figure}

We perform one-pixel attacks on the $ 2 $-class MLP model and CNN model, and generate $ 10,000 $ adversarial perturbations. Although we use the $ 2 $-class model to generate adversarial perturbations, we analyze later that the amplitude distribution of the perturbation generated by $ 9 $-class model is the same as the distribution of $ 2 $-class model. In the differential evolution algorithm, we limit the position of adversarial perturbation to the vicinity of three time samples we select in Section \ref{Section.Loni}. We show the amplitude distribution of these adversarial perturbations near $ 1900 $th time sample and $ 2560 $th time sample in Figures \ref{Fig.MLP-ValueDistr} and \ref{Fig.CNN-ValueDistr}.

Figure \ref{Fig.MLP-ValueDistr} shows the amplitude distribution of adversarial perturbations on MLP model. The most distributed amplitudes of adversarial perturbations are $-5.2$ and $4.8$. Figure \ref{Fig.CNN-ValueDistr} shows the amplitude distribution of the adversarial perturbations on the CNN model. In Figure \ref{Fig.CNN-ValueDistr}, the amplitudes of $ -5 $ and $ 3.8 $ have the most adversarial perturbations. In order to deceive both CNN and MLP, the amplitude that the noise instructions need to generate is within the interval $ \left[ -5.2, -5\right] $  or $ \left[ 3.8, 4.8\right] $, such perturbations are more likely to become \textit{universal perturbation}. The purpose of choosing two amplitude intervals as our criterion for selecting noise instructions is to be able to find more instructions that meet the requirements, and to make these noise instructions effective for various deep learning models.

In experiments, we target AES implementations running over an ARM Cortex-M3 processor. ARM Thumb1 and Thumb2 instructions are treated as a candidate noise instruction set. We capture the energy consumption traces of instructions candidate on the cryptographic device, and select the instruction that can generate a suitable perturbation size as the noise instruction.

The power consumption is not only related to instructions, but also related to operated constants. Generally speaking, \texttt{0xff} causes greater power consumption. In this paper, we chose the four instructions listed in Listing \ref{INI} as our noise instructions. \texttt{r24} in the listing may be any free register, which is determined by the compiler during compilation.

\begin{figure}[htbp]
	\centering 
	\includegraphics[width=8cm]{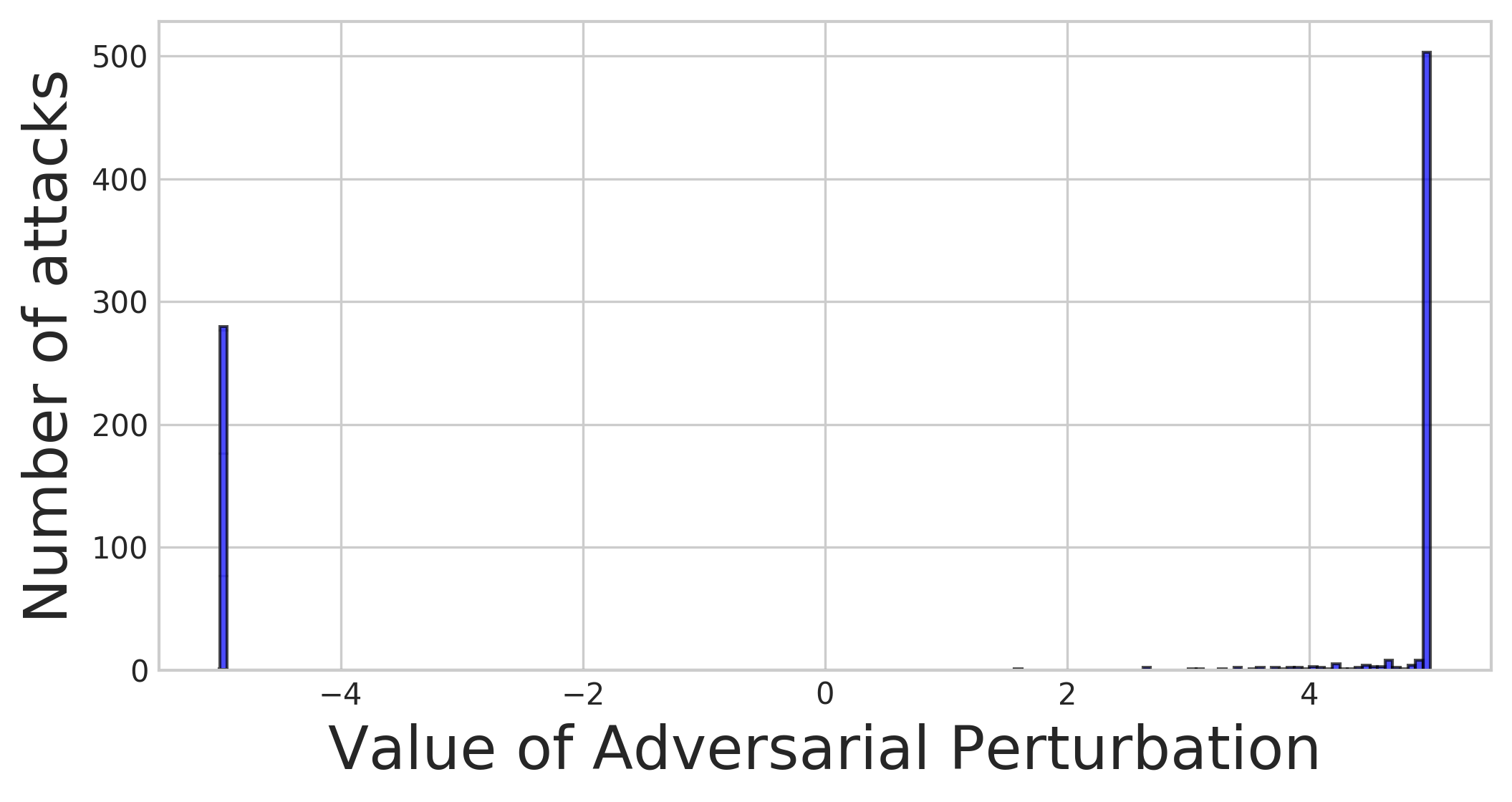}
	\caption{Amplitude distribution of adversarial perturbation near the $ 1900 $th time sample on $ 2 $-class MLP model.} 
	\label{Fig.2classmlp}
\end{figure}

We use $ 9 $-class MLP to generate the adversarial perturbations, and observe their distribution. The results are shown in Figure \ref{Fig.2classmlp}. As with the $ 2 $-class model, the perturbations generated by the $ 9 $-class model is concentrated in the largest and smallest amplitude. The noise instructions which we select using $ 2 $-class model can still generate universal perturbation in $ 9 $-class model.

\begin{center}
	\lstset{
		caption=Inserted noise instructions, 
		basicstyle=\footnotesize, frame=tb,
		xleftmargin=.3\textwidth, xrightmargin=.3\textwidth,
		label=INI
	}
	\begin{lstlisting}
	mov r24, 0xff
	ori r24, 0xff
	ldi r24, 0xff
	in  r24, 0x3d
	\end{lstlisting}
\end{center}

\subsection{Inserting Noise Instructions}
The last step of our countermeasures is to insert the selected noise instructions into the selected positions. In this work, we insert noise instructions into the code at compile time. We start by annotating the assembly file at target positions. The annotated assembly file is recompiled before each invocation of the cryptosystem: when the compiler recognizes these annotations, it randomly picks $ \omega $ noise instructions from Listing \ref{INI}, and inserts them to the code, where $ \omega $ is an integer in $ \left\{ 0, 1, 2\right\}  $. The purpose of inserting different numbers of instructions is to increase the diversity of the code. In order to ensure that the side-channel leakage of each invocation of the cryptosystem becomes an adversarial sample, our approach requires that the cryptographic device recompiles the code at each invocation.

\section{Experimental Evaluation}\label{section-ee}
We evaluate our countermeasures as a defense against deep learning based SCA. In order to demonstrate that our countermeasures are also effective for the classical side-channel attacks, we perform template attacks on our countermeasures. For convenience, we refer to our countermeasures as \textit{one-pixel protection AES} in the following content.
\subsection{Experimental Setup}

We use a STM32F3 board fitted with an Arm Cortex-M3 core running at 32MHz, 16kB of RAM, and 128kB of flash memory. It does not provide any hardware security mechanisms against SCA. Our AES implementation is an unprotected 8-bit implementation that follows the NIST specification.

The side-channel traces are obtained with a Pico5444B PicoScope. The sampling acquisition is performed at $ 96 $ Msample/s. In this scenario, the length of one processor cycle on the side channel trace is three time samples. To ease the temporal alignment of the side-channel traces, a trigger signal is set, and held high during the execution of the first AES round. Using this setup, the security evaluation is performed with stricter conditions than it would be in practice for an adversary.

\subsection{Evaluation Metrics} 
\label{subsectionEM}
We use two metrics to evaluate the performance of different AES implementations against attacks, which are the \textit{rank function} and the \textit{accuracy}. 

With the same previous notations in Section \ref{subsection-nat}, we define the \textit{score} function $S_{M}[k]$ of the key candidates $ k $ :
\begin{equation}
S_{M}[k]=\prod_{i=1}^{M} d_{(i)}[k]. 
\label{score}
\end{equation}
According to Equation \ref{score} we can define the \textit{rank function}:
\begin{equation}
\operatorname{rank}\left(\operatorname{Model}, \mathcal{D}_{\text {profiling}}, \mathcal{D}_{\text {attack}}, M\right)=\left|\left\{k \in \mathcal{K} | \mathrm{S}_{M}[k]>\mathrm{S}_{M}\left[k^{*}\right]\right\}\right|.
\label{rank}
\end{equation}
When the rank of $ k^{*} $ is $ 0 $, we perform a successful key recovery. The larger the $ M $ required to recover the correct key, the better the implementation performs against side-channel attacks. To get a better measure of the rank, it is more suitable to estimate its mean value over several pairs of datasets.

The second metric is the \textit{accuracy} which is commonly used in machine learning. We define it as:
\begin{equation}
\operatorname{acc}\left(\operatorname{Model}, \mathcal{D}_{\text {profiling }}, \mathcal{D}_{\text {attack }}\right)=\frac{\left|\left\{k^{*} \in \mathcal{K} | k^{*}=\operatorname{argmax}_{k \in \mathcal{K}} d_{(i)}[k]\right\}\right|}{\left|\mathcal{D}_{attack}\right|}.
\label{accuracy}
\end{equation}
In this paper, \textit{accuracy} is used to evaluate the performance of $ 2 $-class models. The numbers of elements of each class are equal. Thus, it is adequate using the accuracy as metric. The lower the \textit{accuracy}, the better the security of the countermeasure.

\subsection{Resistance to Practical Attacks}
In this section, we use MLP and CNN to attack three different AES implementations, including unprotected AES implementation, random noise AES and one-pixel protection AES. In order to demonstrate that the effectiveness of one-pixel protection AES is due to the carefully selected insertion position and noise instructions, we implement random noise AES: randomly inserting noise instructions at three random locations in the first round of AES. The inserted noise instructions are randomly selected among instructions that are commonly used in AES programs.

We capture $ 60,000 $ power traces for each AES implementation, and each trace is composed of $ 5,200 $ time samples. To get a better measure of the \textit{rank function} and the \textit{accuracy}, we therefore need to calculate their mean value on several pairs of data sets. Among the $ 60,000 $ traces, $ 10,000 $ traces are randomly selected as the test set, and the remaining $ 50,000 $ traces are used as the training set. Repeating this process $ 10 $ times, we get $ 10 $ different data sets.

In order to analyze the effectiveness of our method on deep learning models with different output classes, we train four deep learning models: $ 9 $-class CNN model (HW leakage model), $ 2 $-class CNN model (LSB leakage model), $ 9 $-class MLP model and $ 2 $-class MLP model. The CNN and MLP architecture used in this paper refers to \cite{prouff2018study}. Since our target are unmasked AES, we reduce the number of epochs.  For MLP, the activation function is Relu and softmax, the optimizer is RMSprop, the learning rate is $ 0.00001 $, batch size is $ 256 $, and the number of epochs is $ 100 $. For CNN, we use the Softmax activation function in the classification layer combined with the Categorical/Binary Cross Entropy loss function. The learning rate is $ 0.0001 $, the optimizer is RMSprop, batch size is $ 256 $, and the number of epochs is $ 10 $.

\subsubsection{2-Class Model}
\begin{figure}[htbp]
	\centering
	\subfigure[Unprotected AES]{
		\includegraphics[width=6.5cm]{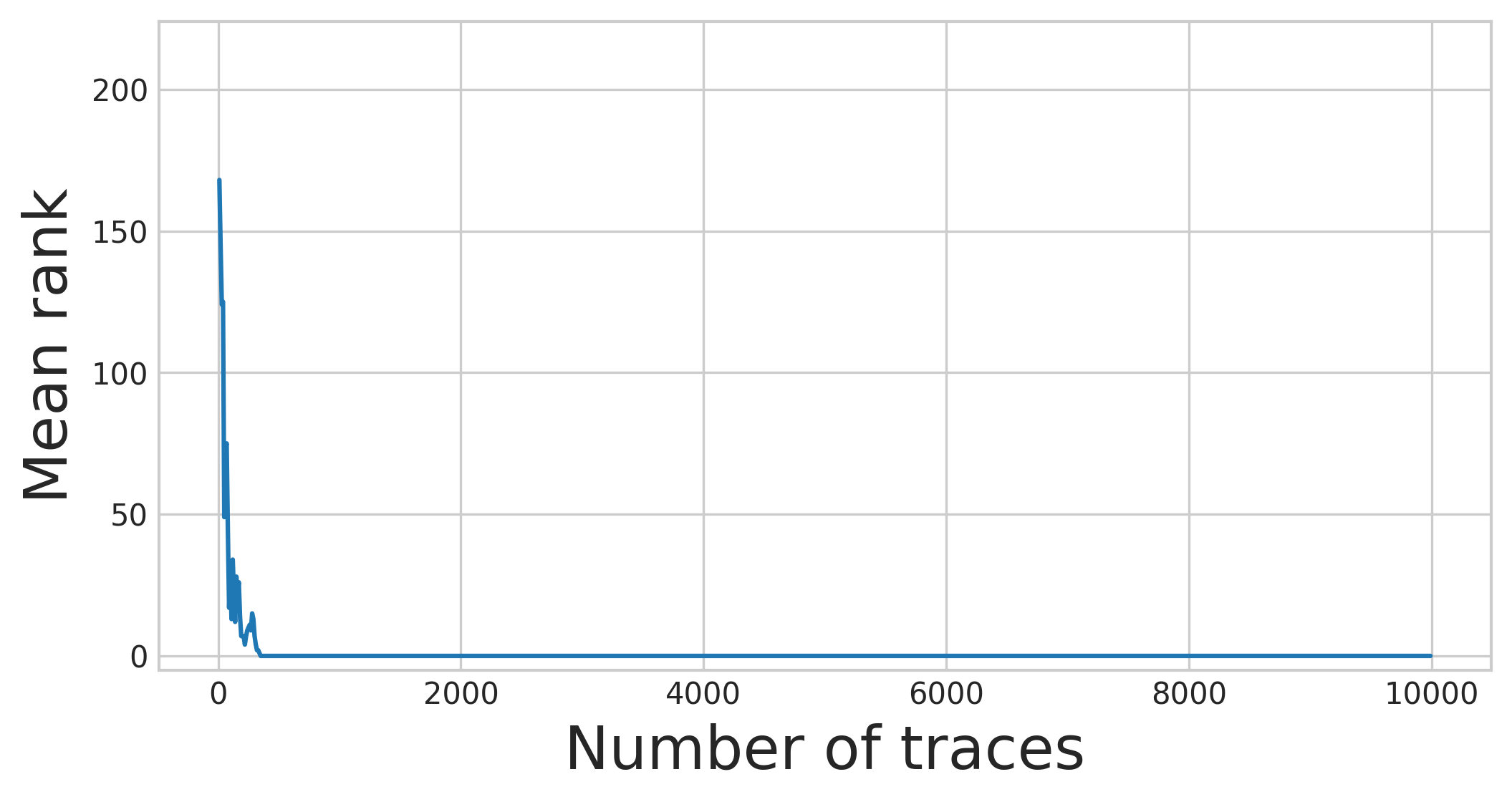}
	}
	\subfigure[Random noise AES]{
		\includegraphics[width=6.5cm]{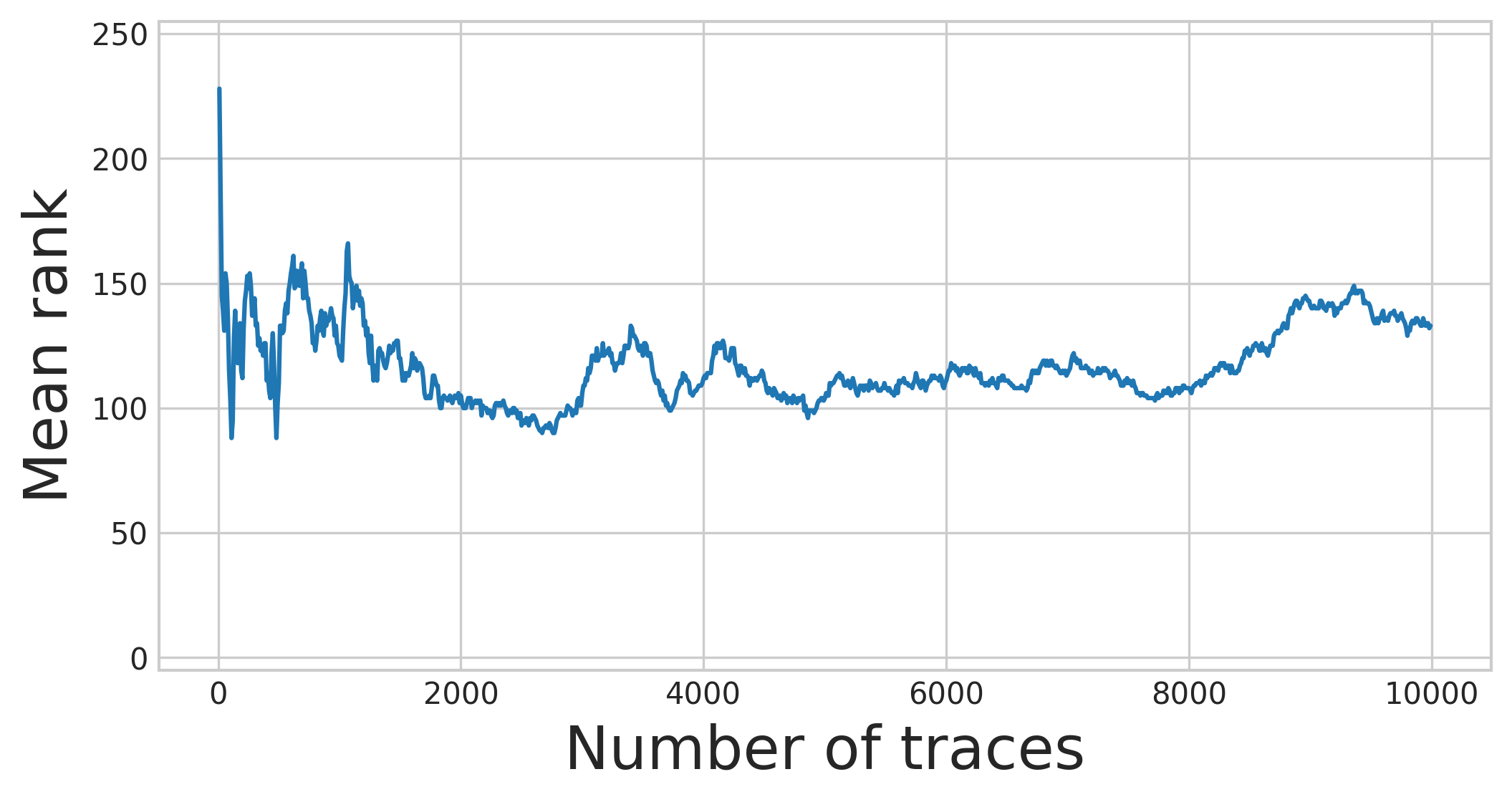}
	}
	\subfigure[One-pixel protection AES]{
		\includegraphics[width=6.5cm]{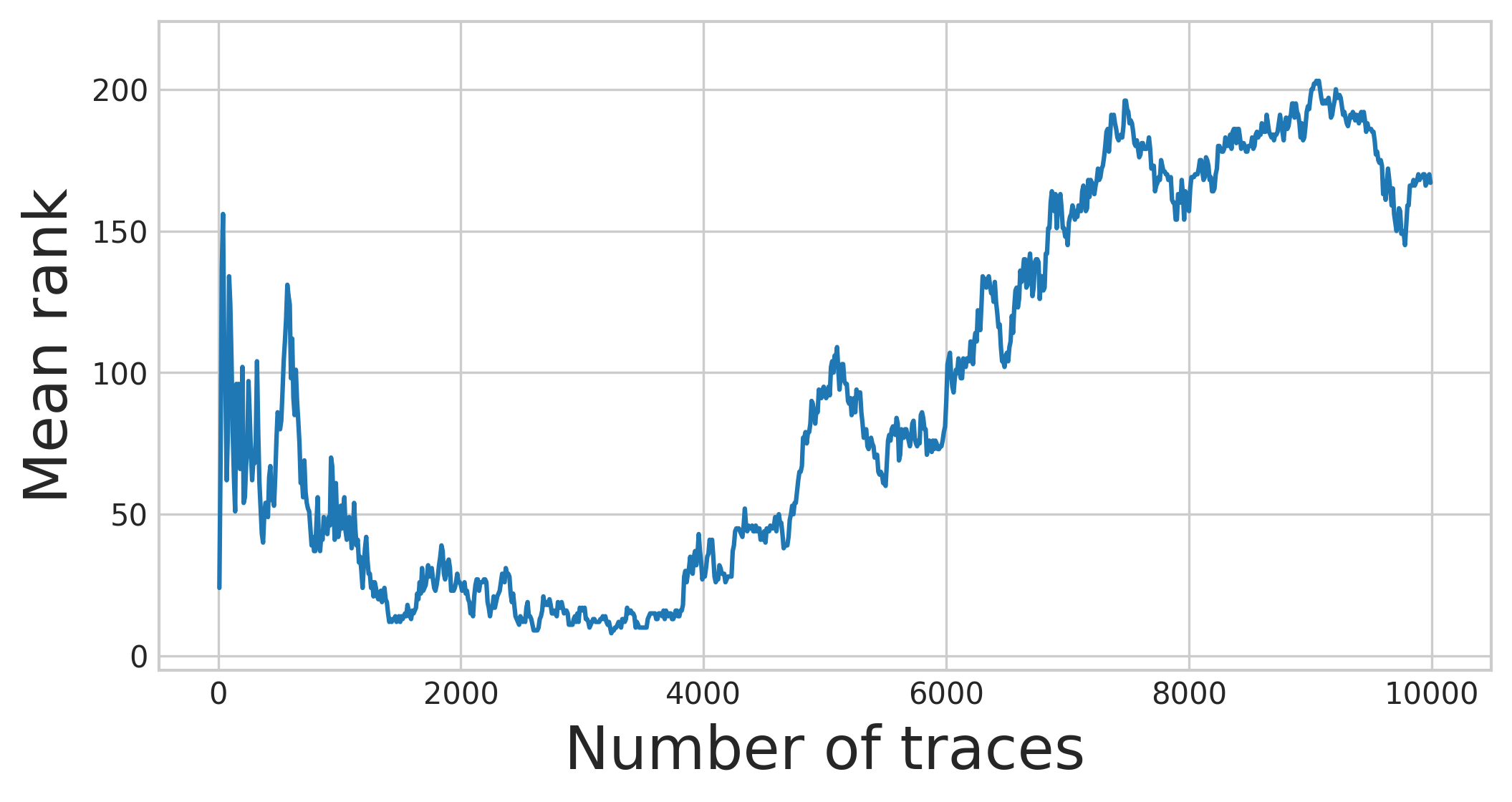}
	}
	\caption{ The mean \textit{rank} of different AES implementations on MLP-based attacks. (a) unprotected AES, $ 340 $ traces are required for a successful key recovery. (b) Random noise AES, MLP cannot retrieve the secret key in $ 10,000 $ traces. (c) One-pixel protection AES, MLP cannot retrieve the secret key in $ 10,000 $ traces. }
	\label{Fig.MLPattack}
\end{figure}

\begin{figure}[htbp]
	\centering
	\subfigure[Unprotected AES]{
		\includegraphics[width=6.5cm]{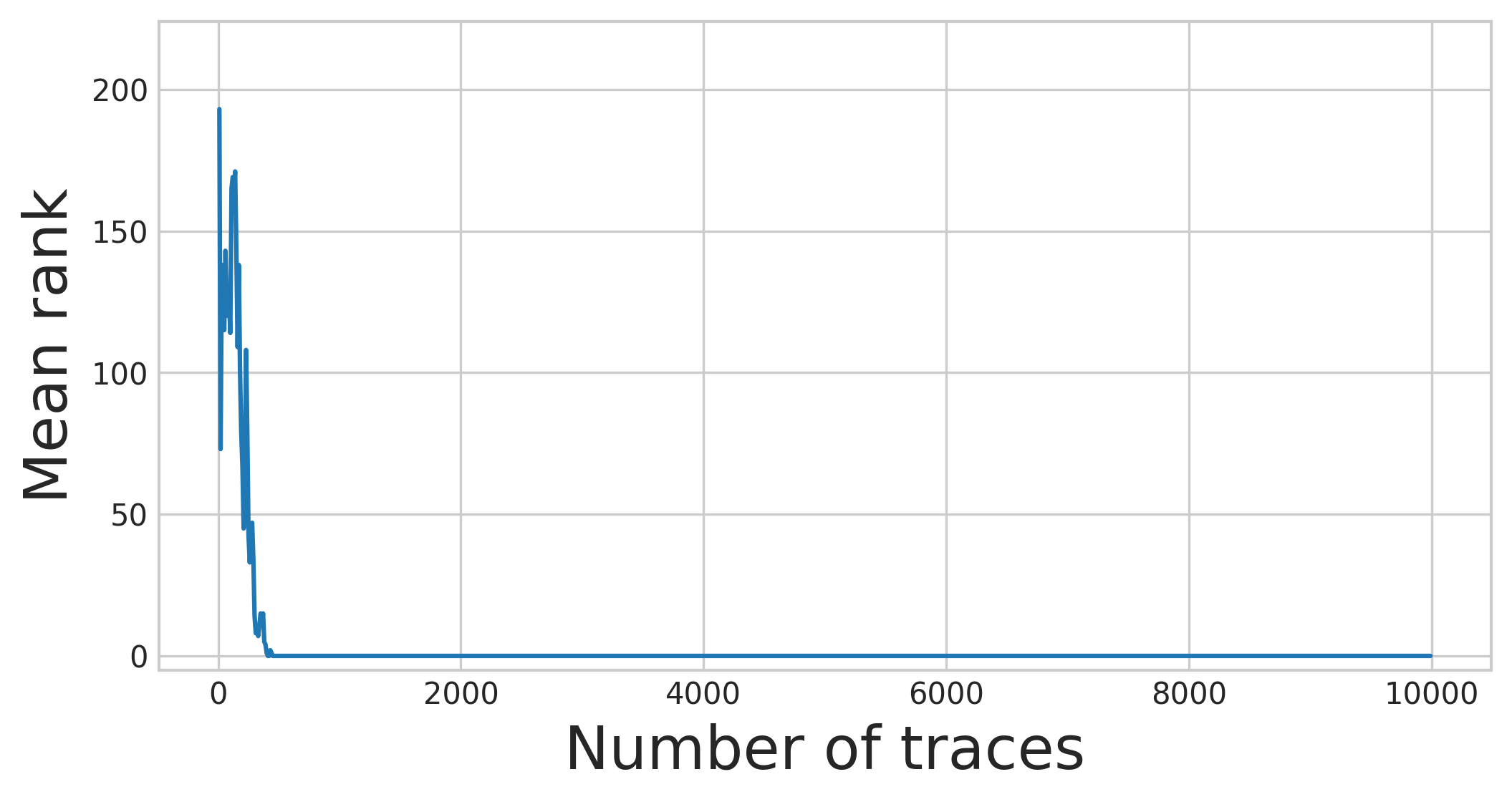}
	}
	\subfigure[Random noise AES]{
		\includegraphics[width=6.5cm]{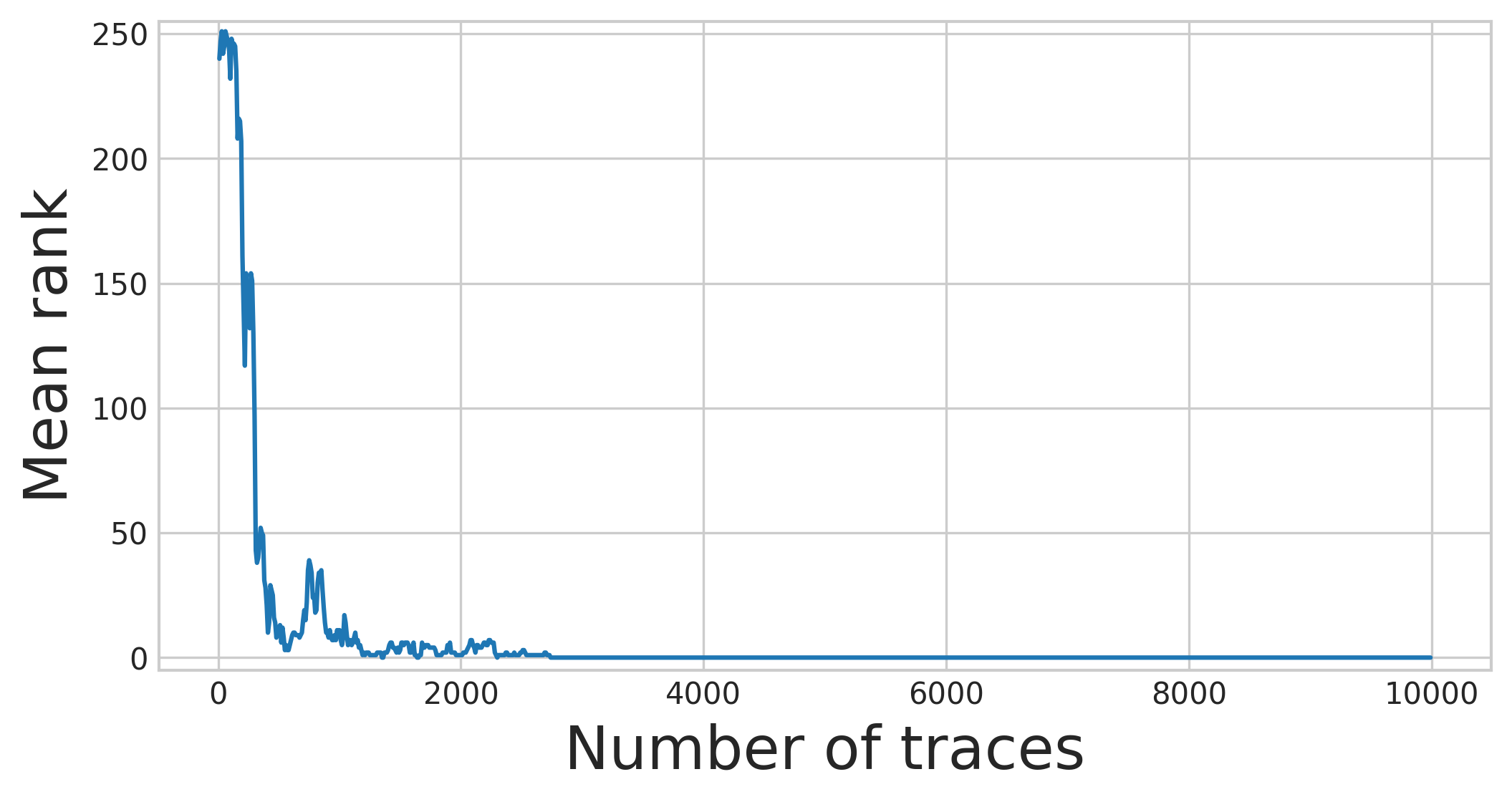}
	}
	\subfigure[One-pixel protection AES]{
		\includegraphics[width=6.5cm]{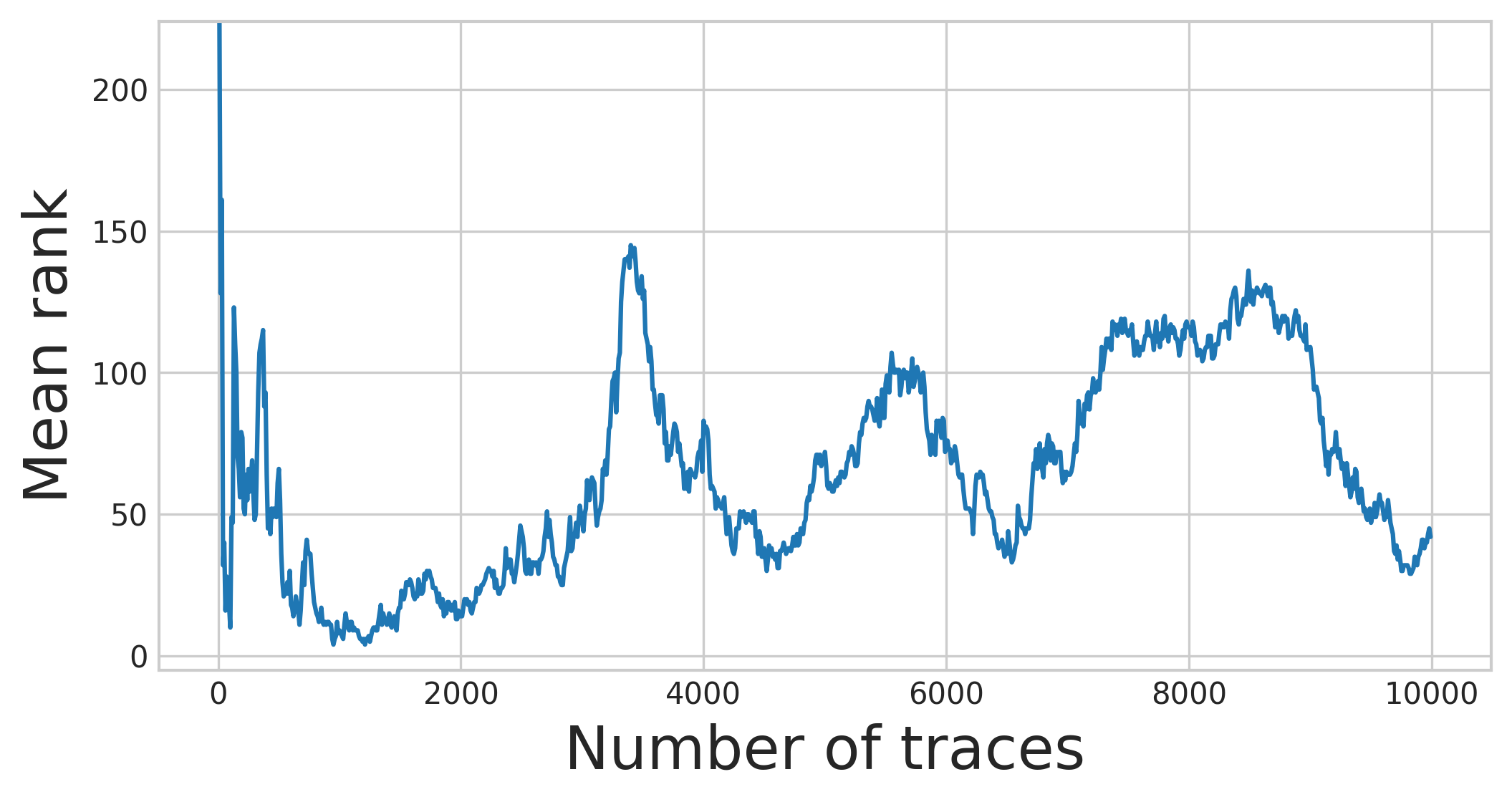}
	}
	\caption{The mean \textit{rank} of different AES implementations on CNN-based attacks. (a) unprotected AES, $ 300 $ traces are required for a successful key recovery. (b) Random-noise AES, $ 2650 $ traces are required for a successful key recovery. (c) One-pixel protection AES, CNN cannot retrieve the secret key in less than $ 10,000 $ traces.}
	\label{Fig.CNNattack}
\end{figure}

Figure \ref{Fig.MLPattack} shows the results of MLP-based attack on three AES implementations. (a), (b), and (c) respectively represent the unprotected AES implementation, the random noise AES and the one-pixel protection AES. Figure \ref{Fig.CNNattack} shows the results of CNN-based attacks on these three AES implementations. For unprotected AES implementation, MLP and CNN recover secret key in $ 340 $ and $ 300 $ traces respectively. For random noise AES, MLP cannot retrieve the key in less than $ 10,000 $ traces, and CNN needs about $ 2650 $ traces to recovery keys. For one-pixel protection AES, MLP and CNN cannot retrieve the secret key in less than $ 10,000 $ traces.

The inserted noise instructions make the power traces of random noise AES into desynchronized traces. Therefore, MLP cannot recover the key of random noise AES. The convolution layer is the main difference between CNN and MLP, and it allows the former has the property of \textit{shift-invariant} \cite{cagli2017convolutional}. Because of this, CNN can still recover the key of random noise AES in presence of desynchronization.

Although CNN can learn shift-invariant features, our countermeasures can still thwart CNN attacks by inserting noise. The noise instructions and insertion positions are carefully selected, and they can generate universal perturbations. Such universal perturbations are doubly universal, both with respect to the data and the network architecture. Therefore, we use a model trained on unprotected AES traces to generate a set of universal perturbations, which can still deceive other models, even when it was trained with different hyperparameters or when it was trained on a different set of traces. 

\begin{figure}[htbp]
	\centering 
	\includegraphics[width=8cm]{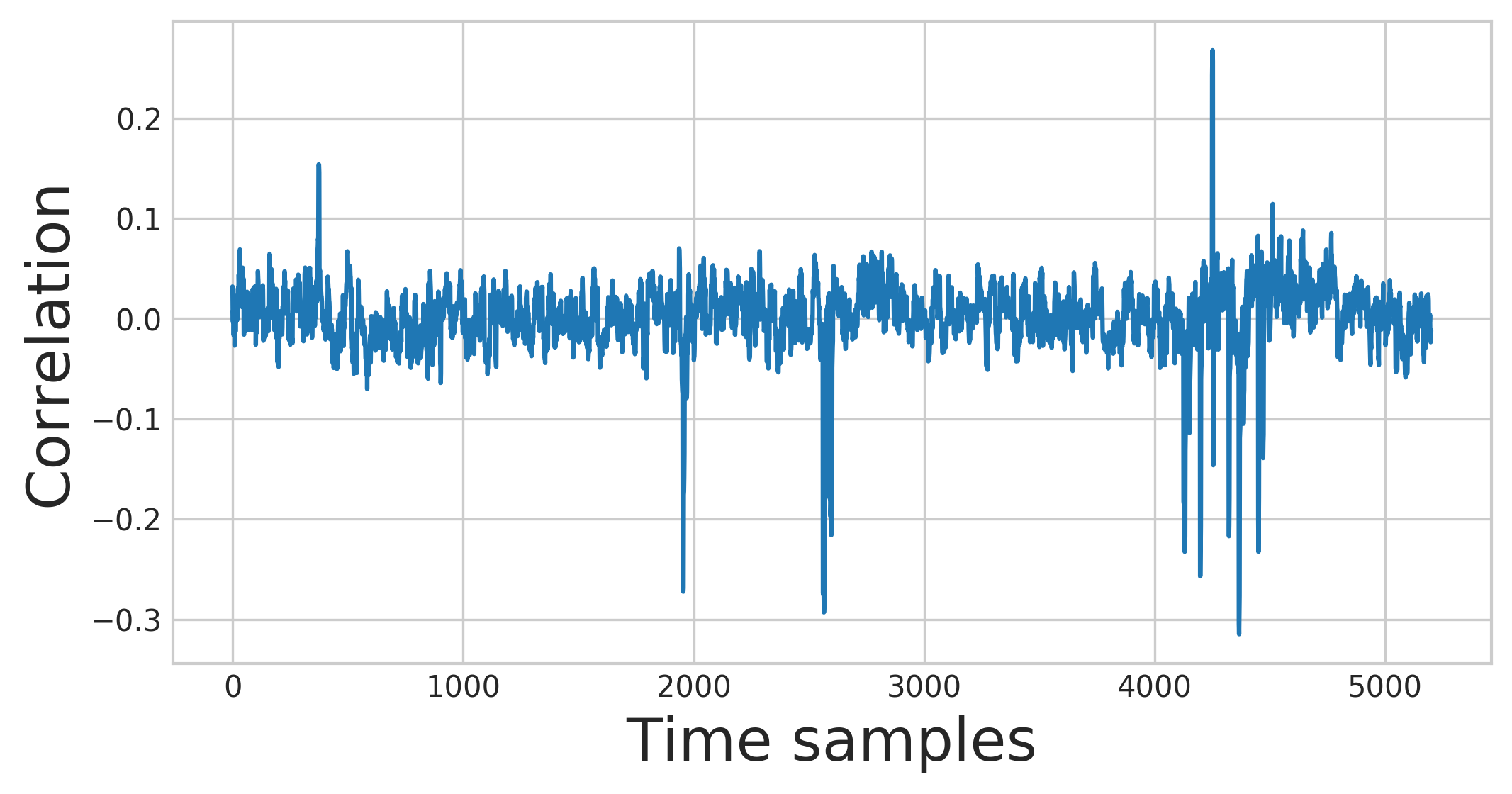}
	\caption{Correlation of side-channel traces} 
	\label{Fig.correlation}
\end{figure}

We calculate correlation factor for the captured side-channel traces (see Figure \ref{Fig.correlation}). The adversary usually selects the points that leak the most information by calculating the correlation factors of the traces. In general, samples with a large correlation leak more side-channel information. Figure \ref{Fig.correlation} shows that the distribution of correlation factors is very similar to the distribution of adversarial perturbations. Therefore, the positions where we insert the noise are the positions with larger correlation. This makes it more difficult for the adversary to recover the key.

Table \ref{Table.accuracy} shows the mean \textit{accuracy} of three AES implementations attacked by deep learning based side-channel attacks. In the experiments in this subsection, the trace is labeled with the LSB of the output of the third Sbox during the first round. So, among the side-channel traces, two classes may be distinguished: $ 0 $ and $ 1 $. When the \textit{accuracy} is closer to $ 0.5 $, it means that the corresponding AES achieves better resistance to side-channel attacks. The table shows that our security protection is very effective, making the accuracy very close to $ 0.5 $. In this situation, the deep learning model can hardly correctly classify the side-channel traces.

\begin{table}[htbp]
	\renewcommand\arraystretch{1.2}
	\centering
	\begin{tabular}{|c|p{2cm}<{\centering}|p{2cm}<{\centering}|p{2cm}<{\centering}|}
		\hline
		\diagbox{Model}{Mean \textit{accuracy}}{AES}&Unprotected &Random-noise &One-pixel protection \\ 
		\hline
		CNN &0.7231&0.6296&0.5063\\
		\hline
		MLP&0.7083&0.5004&0.5023\\
		\hline
	\end{tabular}
	\caption{Mean \textit{accuracy} of different AES implementations attacked by MLP and CNN}
	\label{Table.accuracy}
\end{table}

\subsubsection{9-Class Model}
In the experiments in this subsection, the trace is labeled with the HW of the third output bytes of the Sbox during the first round. To compare with one-pixel protection AES, we also use $ 9 $-class model to attack unprotected AES. As in the previous sections, we have a set of $ 50,000 $ power traces for the profiling phase and have a set of $ 10,000 $ power traces for the attack phase.

\begin{figure}[htbp]
	\centering
	\subfigure[Unprotected AES]{
		\includegraphics[width=6.5cm]{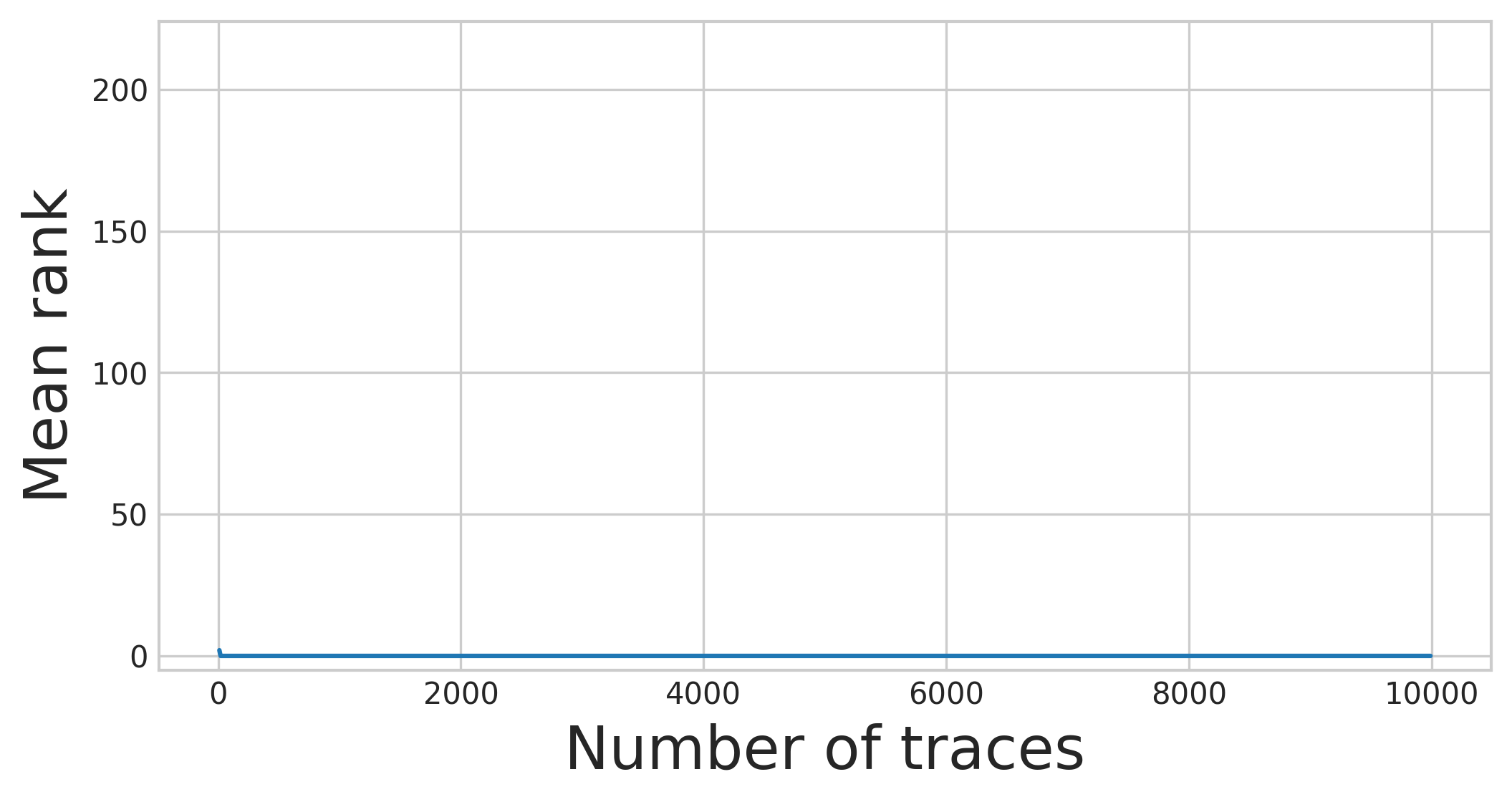}
	}
	\subfigure[One-pixel protection AES]{
		\includegraphics[width=6.5cm]{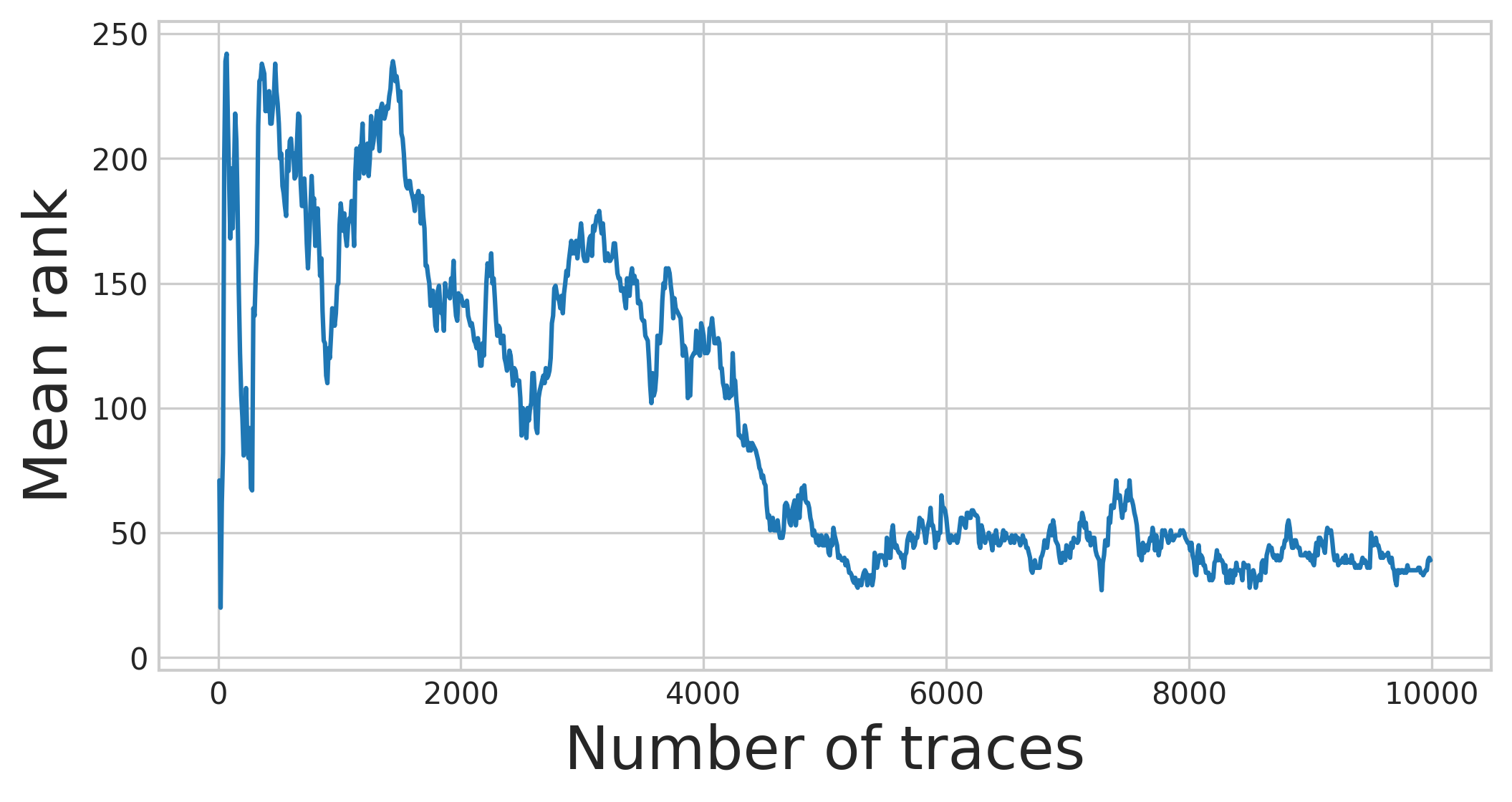}
	}
	\caption{The mean \textit{rank} of different AES implementations on $ 9 $-class MLP attacks. (a) unprotected AES, $ 10 $ traces are required for a successful key recovery. (b) One-pixel protection AES, MLP cannot retrieve the secret key in $ 10,000 $ traces.}
	\label{Fig.MLPattack9class}
\end{figure}

\begin{figure}[htbp]
	\centering
	\subfigure[Unprotected AES]{
		\includegraphics[width=6.5cm]{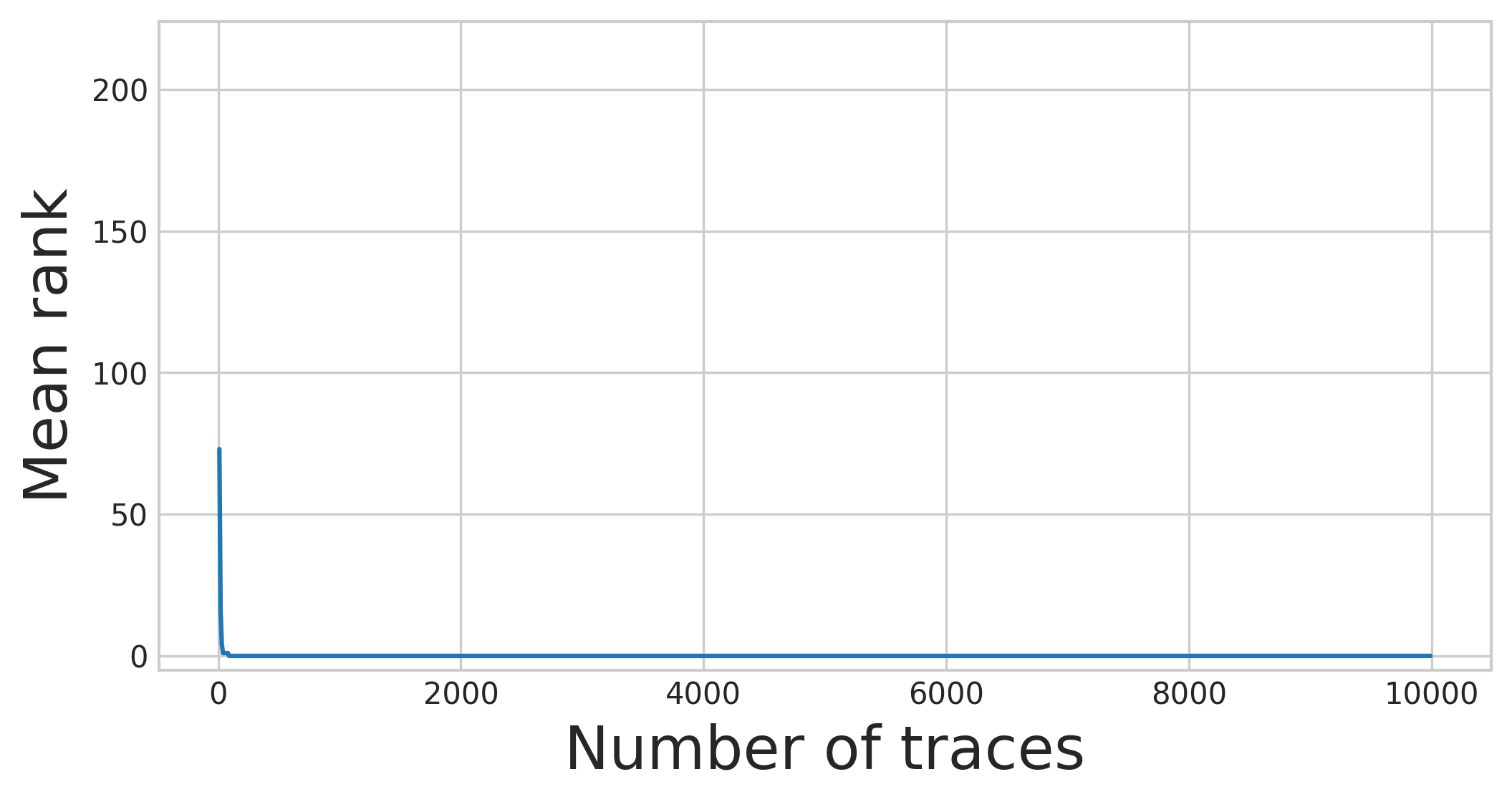}
	}
	\subfigure[One-pixel protection AES]{
		\includegraphics[width=6.5cm]{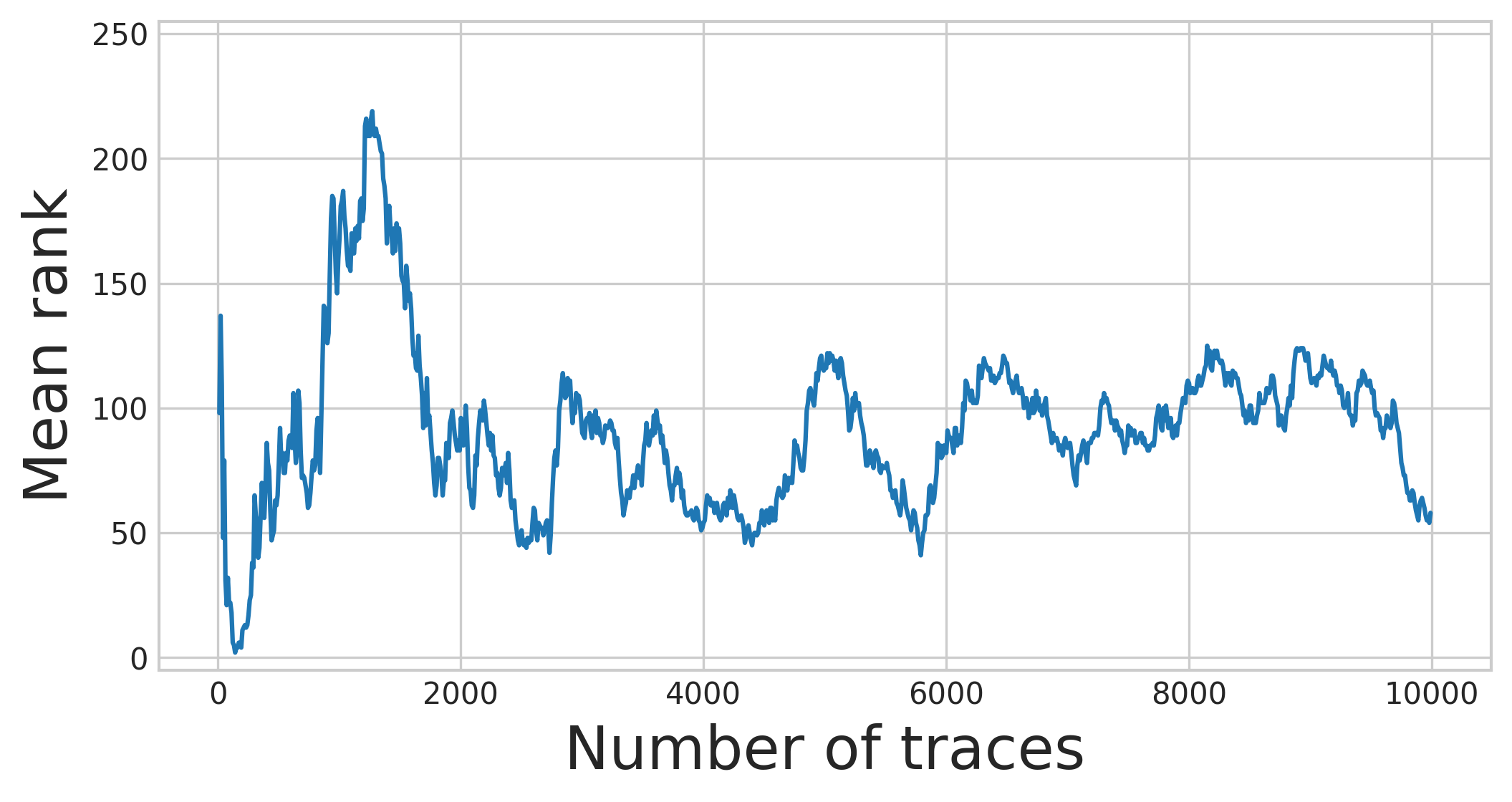}
	}
	\caption{The mean \textit{rank} of different AES implementations on $ 9 $-class CNN attacks. (a) unprotected AES, $ 50 $ traces are required for a successful key recovery. (b) One-pixel protection AES, CNN cannot retrieve the secret key in $ 10,000 $ traces.}
	\label{Fig.CNNPattack9class}
\end{figure}

Figures \ref{Fig.MLPattack9class} and \ref{Fig.CNNPattack9class} show respectively the mean \textit{rank} of different AES implementations on $ 9 $-class MLP attack and CNN attack. The attack results are similar to those of the $ 2 $-class model. One-pixel protection AES can thwart $ 9 $-class CNN and MLP attacks. The reason is that the perturbations generated by inserted noise instructions are universal perturbations, and they can deceive models trained with different hyperparameters.

\subsection{Resistance to Classical Side-Channel Attacks}
\begin{figure}[htbp]
	\centering 
	\includegraphics[width=8cm]{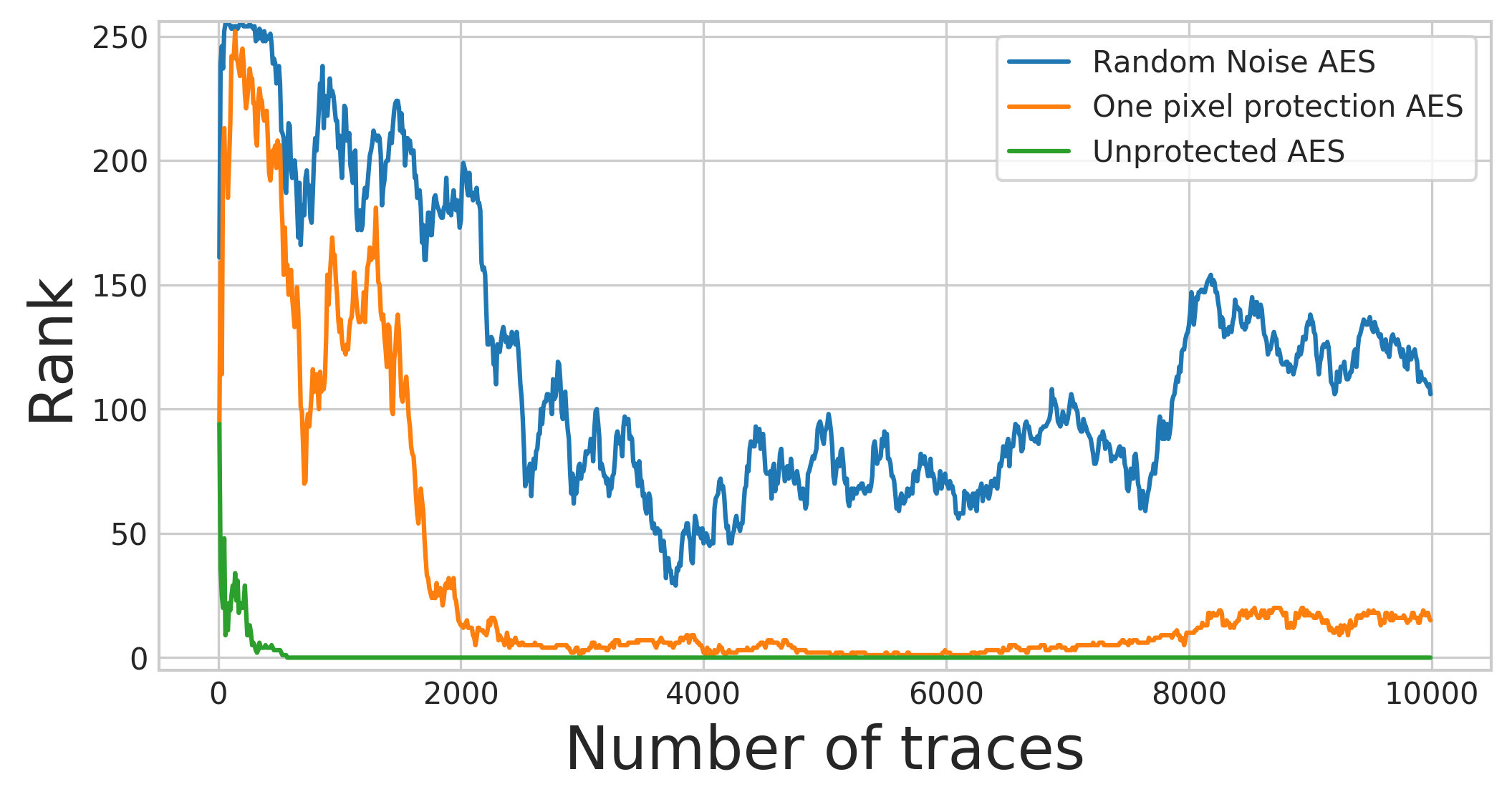}
	\caption{Rank of the correct key attacked by a QDA (Template Attack). For unprotected AES, the key can be retrieved with $ 560 $ traces. For one-pixel protection AES and random noise AES, within $ 10,000 $ traces, the key cannot be retrieved.} 
	\label{Fig.TPA}
\end{figure}

TA are considered as the most successful method in classical SCA. We use TA to evaluate the security level of our countermeasures. We use the TA algorithm described in Section \ref{Section.2.2.2}. We perform template attacks exploiting quadratic discriminant analysis (QDA) which is a well-known generative strategy in the machine learning literature \cite{fisher1936use} to perform classification. We perform QDA on power traces composed of $ 5,200 $ time samples, and do not perform any dimension reduction operation before the TA. As in previous sections, we have a set of $ 50,000 $ power traces for the profiling phase and have a set of $ 10,000 $ power traces for the attack phase. The attack results are illustrated in Figure \ref{Fig.TPA}.

Figure \ref{Fig.TPA} illustrates that, TA only requires $ 560 $ traces to perform a successful key recovery on unprotected AES, but it cannot retrieve the secret key of one-pixel protection AES and random noise AES in less than $ 10,000 $ traces. This demonstrates that, not only for deep learning based SCA, one-pixel protection AES is also effective for classical side-channel attacks. The performance of TA highly depends on some preliminary phases, such as the traces realignment or the selection of the points of interest. Our method can thwart TA for two reasons: 1) we insert noise instructions, and these noise instructions are of variable length, which causes the power traces to be synchronized; 2) the positions of the noise instructions we insert is also the point where the power traces have the greatest correlation, which reduce the correlation between the key $ k $ and the power traces.

\subsection{Execution Time Overhead}

Table \ref{Table.executiontime} compares the execution time (in cycles, measured for $ 1000 $ executions of each AES implementation) of the unprotected AES and our countermeasures. The unprotected AES executes in 5482 processor cycles. The one pixel protection in $ 15952 $ to $ 21328 $ processor cycles (average $ 16418 $ cycles). Table \ref{Table.executiontime} shows that our countermeasures lead to an increase in execution time overhead. The increased execution time overhead is mainly caused by the recompilation at each execution.

\begin{table}[htbp]
	\renewcommand\arraystretch{1.2}
	\centering
	\begin{tabular}{|l|c|c|c|l|c|c|}
		\hline
		\multirow{2}{*}{}                                                                        & \multicolumn{3}{c|}{Unprotected}                                         & \multicolumn{3}{l|}{One-pixel protection}                                   \\ \cline{2-7} 
		& min.                   & avg.                   & max.                   & \multicolumn{1}{c|}{min.} & avg.                   & max.                   \\ \hline
		\multicolumn{1}{|c|}{\begin{tabular}[c]{@{}c@{}}Execution time\\ (cycles)\end{tabular}} & \multicolumn{1}{r|}{5479} & \multicolumn{1}{r|}{5482} & \multicolumn{1}{r|}{5486} & \multicolumn{1}{r|}{15952}    & \multicolumn{1}{r|}{16418} & \multicolumn{1}{r|}{21328} \\ \hline
	\end{tabular}
	\caption{Execution time of the unprotected AES and one-pixel protection AES}
	\label{Table.executiontime}
\end{table}

\section{Discussion and Conclusions}\label{section-con}
We argue that the current preprocessing algorithms cannot break our countermeasures. In order to break countermeasures of inserting noise instructions, some previous works use correlation analysis techniques (e.g. Hidden Markov Models \cite{durvaux2012efficient}) to detect different types of patterns in leakage traces. So that the adversary can distinguish noise instructions and filter them out from the side-channel traces. The reason for the effectiveness of such correlation analysis techniques is that the side-channel profile of noise instructions is different from the profile of useful instructions. Our countermeasures insert noise instructions that are often used in programs, and recompile the code at each invocation. Although this increases the execution time, it ensures that the inserted noise instructions do not have distinguishable headers and tails. Therefore, our countermeasures will not be broken by such techniques. In the context of image classification, there are some works that use dimension reduction techniques to thwart adversarial attacks (e.g. image compression \cite{dziugaite2016study} , Principal Component Analysis \cite{hendrycks2016early}). Moreover, most of the existing dimension reduction are less effective. Their works demonstrate that dimension reduction techniques can reduce the interference of adversarial samples to the model, However, it also reduces the accuracy of the model's classification of normal examples.

Existing methods for generating adversarial examples generally optimize the real examples or add perturbations to the real examples based on the gradient of the model. GAN can train a generator to generate adversarial examples without adding disturbances to specific samples. The samples generated by GAN have the advantage of being more diverse. However, GAN also has the problem that the generated samples change too much compared to the real samples. How to design constraint functions to ensure that GAN can generate adversarial examples of the target category without introducing huge perturbations is the direction of our future work.

In this paper, we present a new direction for achieving protection of cryptographic devices through one-pixel attack techniques. Based on the one-pixel attack techniques, we find the most vulnerable time samples on the side-channel observations, and find the noise instructions that may deceive the deep learning models. We implement our countermeasures and conduct experiments to evaluate the security level. Experiments show that our countermeasures can protect cryptographic devices against deep learning side-channel attacks. Our method is also effective enough against classical side-channel attacks, which makes it more competitive. 

%\section*{Acknowledegment}

\bibliographystyle{alpha}
\bibliography{ref}

\newcommand{\etalchar}[1]{$^{#1}$}
\begin{thebibliography}{MDFFF17}

\bibitem[ABP19]{agosta2019compiler}
Giovanni Agosta, Alessandro Barenghi, and Gerardo Pelosi.
\newblock Compiler-based techniques to secure cryptographic embedded software
  against side channel attacks.
\newblock {\em IEEE Transactions on Computer-Aided Design of Integrated
  Circuits and Systems}, 2019.

\bibitem[AMN{\etalchar{+}}11]{amarilli2011can}
Antoine Amarilli, Sascha M{\"u}ller, David Naccache, Daniel Page, Pablo Rauzy,
  and Michael Tunstall.
\newblock Can code polymorphism limit information leakage?
\newblock In {\em IFIP International Workshop on Information Security Theory
  and Practices}, pages 1--21. Springer, 2011.

\bibitem[ARP07]{ambrose2007rijid}
Jude~Angelo Ambrose, Roshan~G Ragel, and Sri Parameswaran.
\newblock Rijid: random code injection to mask power analysis based side
  channel attacks.
\newblock In {\em Proceedings of the 44th annual Design Automation Conference},
  pages 489--492, 2007.

\bibitem[BBL13]{boulet2013protection}
Frederic Boulet, Michael Barthe, and Thanh-ha Le.
\newblock Protection of applets against hidden-channel analyses, November~21
  2013.
\newblock US Patent App. 13/997,136.

\bibitem[BCHC18]{belleville2018automated}
Nicolas Belleville, Damien Courouss{\'e}, Karine Heydemann, and Henri-Pierre
  Charles.
\newblock Automated software protection for the masses against side-channel
  attacks.
\newblock {\em ACM Transactions on Architecture and Code Optimization (TACO)},
  15(4):47, 2018.

\bibitem[CBR{\etalchar{+}}16]{courousse2016runtime}
Damien Courouss{\'e}, Thierno Barry, Bruno Robisson, Philippe Jaillon, Olivier
  Potin, and Jean-Louis Lanet.
\newblock Runtime code polymorphism as a protection against side channel
  attacks.
\newblock In {\em IFIP International Conference on Information Security Theory
  and Practice}, pages 136--152. Springer, 2016.

\bibitem[CDP17]{cagli2017convolutional}
Eleonora Cagli, C{\'e}cile Dumas, and Emmanuel Prouff.
\newblock Convolutional neural networks with data augmentation against
  jitter-based countermeasures.
\newblock In {\em International Conference on Cryptographic Hardware and
  Embedded Systems}, pages 45--68. Springer, 2017.

\bibitem[CK10]{coron2010analysis}
Jean-S{\'e}bastien Coron and Ilya Kizhvatov.
\newblock Analysis and improvement of the random delay countermeasure of ches
  2009.
\newblock In {\em International Workshop on Cryptographic Hardware and Embedded
  Systems}, pages 95--109. Springer, 2010.

\bibitem[CRR02]{chari2002template}
Suresh Chari, Josyula~R Rao, and Pankaj Rohatgi.
\newblock Template attacks.
\newblock In {\em International Workshop on Cryptographic Hardware and Embedded
  Systems}, pages 13--28. Springer, 2002.

\bibitem[DGR16]{dziugaite2016study}
Gintare~Karolina Dziugaite, Zoubin Ghahramani, and Daniel~M Roy.
\newblock A study of the effect of jpg compression on adversarial images.
\newblock {\em arXiv preprint arXiv:1608.00853}, 2016.

\bibitem[DRS{\etalchar{+}}12]{durvaux2012efficient}
Fran{\c{c}}ois Durvaux, Mathieu Renauld, Fran{\c{c}}ois-Xavier Standaert, Loic
  van~Oldeneel tot Oldenzeel, and Nicolas Veyrat-Charvillon.
\newblock Efficient removal of random delays from embedded software
  implementations using hidden markov models.
\newblock In {\em International Conference on Smart Card Research and Advanced
  Applications}, pages 123--140. Springer, 2012.

\bibitem[DS10]{das2010differential}
Swagatam Das and Ponnuthurai~Nagaratnam Suganthan.
\newblock Differential evolution: A survey of the state-of-the-art.
\newblock {\em IEEE transactions on evolutionary computation}, 15(1):4--31,
  2010.

\bibitem[Fis36]{fisher1936use}
Ronald~A Fisher.
\newblock The use of multiple measurements in taxonomic problems.
\newblock {\em Annals of eugenics}, 7(2):179--188, 1936.

\bibitem[GMO01]{gandolfi2001electromagnetic}
Karine Gandolfi, Christophe Mourtel, and Francis Olivier.
\newblock Electromagnetic analysis: Concrete results.
\newblock In {\em International workshop on cryptographic hardware and embedded
  systems}, pages 251--261. Springer, 2001.

\bibitem[HG16]{hendrycks2016early}
Dan Hendrycks and Kevin Gimpel.
\newblock Early methods for detecting adversarial images.
\newblock {\em arXiv preprint arXiv:1608.00530}, 2016.

\bibitem[KGB16]{kurakin2016adversarial}
Alexey Kurakin, Ian Goodfellow, and Samy Bengio.
\newblock Adversarial examples in the physical world.
\newblock {\em arXiv preprint arXiv:1607.02533}, 2016.

\bibitem[KJJ99]{kocher1999differential}
Paul Kocher, Joshua Jaffe, and Benjamin Jun.
\newblock Differential power analysis.
\newblock In {\em Annual International Cryptology Conference}, pages 388--397.
  Springer, 1999.

\bibitem[Koc96]{kocher1996timing}
Paul~C Kocher.
\newblock Timing attacks on implementations of diffie-hellman, rsa, dss, and
  other systems.
\newblock In {\em Annual International Cryptology Conference}, pages 104--113.
  Springer, 1996.

\bibitem[KPH{\etalchar{+}}19]{kim2019make}
Jaehun Kim, Stjepan Picek, Annelie Heuser, Shivam Bhasin, and Alan Hanjalic.
\newblock Make some noise. unleashing the power of convolutional neural
  networks for profiled side-channel analysis.
\newblock {\em IACR Transactions on Cryptographic Hardware and Embedded
  Systems}, pages 148--179, 2019.

\bibitem[MDFFF17]{moosavi2017universal}
Seyed-Mohsen Moosavi-Dezfooli, Alhussein Fawzi, Omar Fawzi, and Pascal
  Frossard.
\newblock Universal adversarial perturbations.
\newblock In {\em Proceedings of the IEEE conference on computer vision and
  pattern recognition}, pages 1765--1773, 2017.

\bibitem[MDM16]{martinasek2016profiling}
Zdenek Martinasek, Petr Dzurenda, and Lukas Malina.
\newblock Profiling power analysis attack based on mlp in dpa contest v4. 2.
\newblock In {\em 2016 39th International Conference on Telecommunications and
  Signal Processing (TSP)}, pages 223--226. IEEE, 2016.

\bibitem[MDP20]{masure2020comprehensive}
Lo{\"\i}c Masure, C{\'e}cile Dumas, and Emmanuel Prouff.
\newblock A comprehensive study of deep learning for side-channel analysis.
\newblock {\em IACR Transactions on Cryptographic Hardware and Embedded
  Systems}, pages 348--375, 2020.

\bibitem[MHM13]{martinasek2013optimization}
Zdenek Martinasek, Jan Hajny, and Lukas Malina.
\newblock Optimization of power analysis using neural network.
\newblock In {\em International Conference on Smart Card Research and Advanced
  Applications}, pages 94--107. Springer, 2013.

\bibitem[MMS{\etalchar{+}}18]{madry2018towards}
Aleksander Madry, Aleksandar Makelov, Ludwig Schmidt, Dimitris Tsipras, and
  Adrian Vladu.
\newblock Towards deep learning models resistant to adversarial attacks.
\newblock In {\em International Conference on Learning Representations}, 2018.

\bibitem[MMT15]{martinasek2015profiling}
Zdenek Martinasek, Lukas Malina, and Krisztina Trasy.
\newblock Profiling power analysis attack based on multi-layer perceptron
  network.
\newblock In {\em Computational Problems in Science and Engineering}, pages
  317--339. Springer, 2015.

\bibitem[MPP16]{maghrebi2016breaking}
Houssem Maghrebi, Thibault Portigliatti, and Emmanuel Prouff.
\newblock Breaking cryptographic implementations using deep learning
  techniques.
\newblock In {\em International Conference on Security, Privacy, and Applied
  Cryptography Engineering}, pages 3--26. Springer, 2016.

\bibitem[PEvW18]{perin2018lowering}
Guilherme Perin, Baris Ege, and Jasper van Woudenberg.
\newblock Lowering the bar: Deep learning for side channel analysis.
\newblock 2018.

\bibitem[PJB19]{picek2019poster}
Stjepan Picek, Dirmanto Jap, and Shivam Bhasin.
\newblock Poster: When adversary becomes the guardian--towards side-channel
  security with adversarial attacks.
\newblock In {\em Proceedings of the 2019 ACM SIGSAC Conference on Computer and
  Communications Security}, pages 2673--2675, 2019.

\bibitem[PSB{\etalchar{+}}18]{prouff2018study}
Emmanuel Prouff, Remi Strullu, Ryad Benadjila, Eleonora Cagli, and Cecile
  Dumas.
\newblock Study of deep learning techniques for side-channel analysis and
  introduction to ascad database.
\newblock {\em IACR Cryptology ePrint Archive}, 2018:53, 2018.

\bibitem[SKM{\etalchar{+}}18]{singh2018exploiting}
Arvind Singh, Monodeep Kar, Sanu Mathew, Anand Rajan, Vivek De, and Saibal
  Mukhopadhyay.
\newblock Exploiting on-chip power management for side-channel security.
\newblock In {\em 2018 Design, Automation \& Test in Europe Conference \&
  Exhibition (DATE)}, pages 401--406. IEEE, 2018.

\bibitem[SP97]{storn1997differential}
Rainer Storn and Kenneth Price.
\newblock Differential evolution--a simple and efficient heuristic for global
  optimization over continuous spaces.
\newblock {\em Journal of global optimization}, 11(4):341--359, 1997.

\bibitem[SVS19]{su2019one}
Jiawei Su, Danilo~Vasconcellos Vargas, and Kouichi Sakurai.
\newblock One pixel attack for fooling deep neural networks.
\newblock {\em IEEE Transactions on Evolutionary Computation}, 2019.

\bibitem[SZS{\etalchar{+}}13]{szegedy2013intriguing}
Christian Szegedy, Wojciech Zaremba, Ilya Sutskever, Joan Bruna, Dumitru Erhan,
  Ian Goodfellow, and Rob Fergus.
\newblock Intriguing properties of neural networks.
\newblock {\em arXiv preprint arXiv:1312.6199}, 2013.

\bibitem[Tim19]{timon2019non}
Benjamin Timon.
\newblock Non-profiled deep learning-based side-channel attacks with
  sensitivity analysis.
\newblock {\em IACR Transactions on Cryptographic Hardware and Embedded
  Systems}, pages 107--131, 2019.

\bibitem[YSG{\etalchar{+}}19]{yan2019attack}
Mengjia Yan, Read Sprabery, Bhargava Gopireddy, Christopher Fletcher, Roy
  Campbell, and Josep Torrellas.
\newblock Attack directories, not caches: Side channel attacks in a
  non-inclusive world.
\newblock In {\em 2019 IEEE Symposium on Security and Privacy (SP)}, pages
  888--904. IEEE, 2019.

\end{thebibliography}

\end{document}